\documentclass[pra,twocolumn,preprintnumbers,amsmath,amssymb,nofootinbib]{revtex4}
\usepackage{amsmath}
\usepackage{amssymb}
\usepackage{slashed}
\usepackage{graphicx}
\usepackage{graphics}
\usepackage{epsfig}
\usepackage{bbm,bm}
\usepackage{psfrag}
\usepackage{hyperref}

\usepackage{color}
\definecolor{comment}{rgb}{0.9,0,0}

\newcommand{\dslash}{\partial \!\!\!/}

\newcommand{\pslash}{p \hspace{-0.15cm}/}

\newcommand{\be}{\begin{eqnarray}}
\newcommand{\ee}{\end{eqnarray}}

\newcommand{\Nf}{N_{\text{f}}}
\newcommand{\NL}{N_{\text{L}}}

\newcommand{\pat}{\partial_t}
\newcommand{\Eqref}[1]{Eq.~\eqref{#1}}

\begin{document}

\title{Towards an Asymptotic-Safety Scenario for Chiral Yukawa Systems}
%\date{\today}
\author{Holger Gies, Stefan Rechenberger and Michael M. Scherer}


\affiliation{\mbox{\it Theoretisch-Physikalisches Institut, Friedrich-Schiller-Universit{\"a}t Jena,}
\mbox{\it Max-Wien-Platz 1, D-07743 Jena, Germany}
\mbox{\it E-mail: { holger.gies@uni-jena.de, michael.scherer@uni-jena.de,}}
\mbox{\it {stefan.rechenberger@uni-jena.de}}
}

\begin{abstract} 
 We search for asymptotic safety in a Yukawa system with a chiral
  U$(N_\mathrm{L})_\mathrm{L}\otimes$U$(1)_\mathrm{R}$ symmetry, serving as a
  toy model for the standard-model Higgs sector. Using the functional RG as a
  nonperturbative tool, the leading-order derivative expansion exhibits admissible
  non-Gau\ss ian fixed-points for $1 \leq N_{\mathrm{L}} \leq 57$
  which arise from a conformal threshold behavior induced by self-balanced
  boson-fermion fluctuations. If present in the full theory, the fixed-point
  would solve the triviality problem. Moreover, as one fixed point has only
  one relevant direction even with a reduced hierarchy problem, the Higgs mass
  as well as the top mass are a prediction of the theory in terms of the Higgs
  vacuum expectation value. In our toy model, the fixed point is destabilized
  at higher order due to massless Goldstone and fermion fluctuations, which
  are particular to our model and have no analogue in the standard model.
\end{abstract}

\maketitle

\section{Introduction}

The Higgs sector is a crucial building block of the standard model of particle
physics and parameterizes the masses of matter fields and weak gauge
bosons. This successful parameterization goes along with two problems of the
standard model which give rise to the belief that the standard model should be
embedded in a larger fundamental framework: the hierarchy problem and the
triviality problem. These problems have initiated many further developments.

The triviality problem renders the theory ill-defined from a fundamental point
of view, since it inhibits an extension of the standard model to arbitrarily
high momentum scales
\cite{Wilson:1973jj,Luscher:1987ek,Hasenfratz:1987eh,Heller:1992js,%
  Callaway:1988ya,Rosten:2008ts}. The scale of the maximum ultra-violet (UV)
extension $\Lambda_{\text{UV,max}}$ induced by triviality is expected to be
related to the Landau pole of perturbation theory. Of course, the Landau pole,
i.e., the divergence of a perturbative running coupling at a finite UV scale,
in the first place signals the breakdown of perturbation theory. Near the
Landau pole, nonperturbative physics can set in and severely modify the
picture. So, a study of the triviality problem and the existence of a finite
$\Lambda_{\text{UV,max}}$ therefore requires a nonperturbative tool. This has
also inspired a number of lattice investigations of scalar and chiral Yukawa
systems \cite{Smit:1989tz,Lee:1989mi,Kuti:1987nr,Jansen:1993jj}. 

 Whereas the triviality problem is a true conceptual problem of the standard
 model, the hierarchy problem, i.e., the possible existence of a huge
 difference between the electroweak scale and an underlying fundamental scale
 such as the GUT or the Planck scale, is only a problem of unnaturally
 fine-tuned initial conditions (in this case of the mass parameter in the
 Higgs potential at the underlying scale).

Solutions to these two problems are often sought by introducing field theories
with new degrees of freedom or higher symmetries or by going beyond quantum
field theory. The guiding principle of this work is more conservative, as we
intend to identify (or rule out) possible solutions within quantum field theory
essentially using the same or very similar degrees of freedom of the standard
model. Quantum field theory indeed offers a framework for such solutions in
terms of Weinberg's asymptotic safety scenario
\cite{Weinberg:1976xy,Percacci:2007sz,Weinberg:2009}, which has already been
investigated in a variety of models ranging from four-fermion models
\cite{Rosenstein:pt,Gies:2003dp,Schwindt:2008gj}, simple Yukawa systems
\cite{GiesScherer:2009}, nonlinear sigma models in $d>2$
\cite{Codello:2008qq}, and extra-dimensional gauge theories \cite{Gies:2003ic}
to gravity \cite{Reuter:1996cp,Percacci:2003jz}. For the asymptotic-safety
scenario to apply, a fixed point of the renormalization group (RG) flow in the
space of couplings has to exist. If the system sits on an RG trajectory that
hits the fixed point in the UV, the UV cutoff can safely be taken to infinity,
and the theory can remain valid to arbitrarily short distance scales.

In fact, a suitable fixed point has recently been identified in a
Z${}_2$-symmetric Yukawa system \cite{GiesScherer:2009}, serving as a simple
toy model for the top-Higgs sector of the standard model. Moreover, a generic
mechanism inducing such a fixed point has been proposed which relies on a
conformal behavior of the Higgs vacuum expectation value (vev). This
conformal-vev mechanism is generated by a dynamical self-balancing of bosonic
and fermionic degrees of freedom in the UV. In the simple Z$_{2}$ model, this
balancing actually occurs only for small fractional flavor number $\Nf\lesssim
0.3$. This observation is a strong motivation to consider more realistic
Yukawa systems and to explore the potential of the conformal-vev mechanism for
the UV problems of the standard model. Another scenario to circumvent the
problem of triviality in such a Z${}_2$-symmetric Yukawa system in the same
framework occurs by coupling the system to a gravitational background
\cite{percacci:2009}.

This work is devoted to an investigation of a chiral
$U(N_\mathrm{L})_\mathrm{L}\otimes U(1)_\mathrm{R}$ Yukawa model, serving as a
more sophisticated toy-model for the Higgs sector of the standard model (or of
a GUT-like theory). The model is designed to have a left-handed chiral sector
as is typical for the Higgs sector of the standard model. The number of
left-handed fermions $\NL$ is left as a free parameter in order to study the
dependence of a potential fixed point on the varying numbers of degrees of
freedom. As will be demonstrated by our analysis, this chiral structure
facilitates a more boson-dominated Higgs vacuum expectation value which is a
prerequisite for conformal-vev mechanism to work. Using the functional RG as a
nonperturbative approach, we systematically search for the existence of
non-Gau\ss ian interacting fixed points of the RG flow which could allow for
an extension of the model to arbitrarily high momentum scales and render the
system asymptotically safe.

A particularly attractive by-product of an asymptotically safe Higgs sector is
given by the fact that the number of physical parameters is dictated by the
number of RG-relevant directions at the fixed point and is thus an inherent
property of the model. As this number can actually be smaller than the
corresponding one at the Gau\ss ian fixed point (defining the ``perturbative''
standard model), asymptotic safety can lead to a reduction of physical
parameters and hence have more predictive power. For instance, the interacting
fixed point in the Z${}_2$-invariant toy model has one relevant direction less
than the perturbative fixed point. As an immediate consequence, the value of
the Higgs mass becomes a prediction once the Higgs vev and the top mass are
fixed. For a fixed point of our chiral model discussed in this work, even the
top mass can become a prediction, demonstrating the predictive and
constraining power of the asymptotic safety scenario. 

A crucial question for all nonperturbative techniques is the systematic
consistency and reliability of the results. In this work, we compute the RG
flow of the model in a systematic derivative expansion of the effective
action. This expansion is controlled if the momentum dependence of full
effective vertices takes only little influence on the flow. A direct means for
measuring this influence is the size of the anomalous dimensions $\eta$ of the
fields, since next-to-leading order contributions couple to the leading-order
derivative expansion only via terms $\sim\eta$. Monitoring the size of $\eta$
thus gives us a direct estimate of the reliability of our results.  Whereas
the anomalous dimensions at the fixed point of the Z${}_2$ model were indeed
found to be small, the anomalous dimensions at the fixed point of the present
model can become large. Hence, the results within the present model have to be
taken with a grain of salt. The reason for the difference between the two
models lies in the existence of Goldstone bosons as well as massless fermions
in the present model which contribute dominantly to the anomalous
dimensions. As these massless modes are not present in the standard model, we
expect that the leading-order derivative expansion (where $\eta=0$) of the
present system can serve as a model for the Higgs-Yukawa sector of a more
realistic gauged version. From this viewpoint, the essential idea to have a
non-Gau\ss ian fixed point works and provides us with a highly predictive
theory.

This paper is organized as follows: in Sect.~\ref{sec:as}, we summarize the
essence of the conformal-vev mechanism and introduce the concept of asymptotic
safety. Sect.~\ref{sec:RG} discusses the nonperturbative construction of the
effective action in terms of the functional RG. The fact that the resulting
flow equations do not support asymptotic safety in the symmetric regime is
briefly elucidated in Sect.~\ref{sec:SYM}. In Sect.~\ref{sec:SSB}, the regime
of spontaneous symmetry breaking is explored to leading order, revealing
non-Gau\ss ian fixed points with physically appealing properties. The
resulting predictive power of the asymptotic-safety scenario is described in
Sect.~\ref{sec:PP}. Sect.~\ref{sec:NLO} summarizes the problems of the present
toy-model occurring at next-to-leading order in the derivative
expansion. Conclusions are drawn in Sect.~\ref{sec:conc}. Technical details on
the derivation of the flow equations are given in the appendices.

\section{An asymptotic-safety scenario for Yukawa Systems}
\label{sec:as}

\subsection{Conformal vacuum expectation value}

The model building of the present work is strongly motivated by qualitative considerations about the loop contributions to the running of the dimensionless version of the bosonic field expectation value $v$ in the regime of spontaneous symmetry breaking (SSB) where $v>0$. Our central idea is that the contributions with opposite sign from bosonic and fermionic fluctuations to
the vev can be balanced such that the vev exhibits a conformal behavior, $v\equiv\langle\varphi\rangle\sim
k$. Here, $k$ is the scale at which we consider the couplings of the
system. The dimensionless squared vev $\kappa= \frac{1}{2}v^2/k^2$ has a flow equation of
the form
\begin{equation}
\pat \kappa \equiv \pat \frac{v^2}{2 k^2} = -2 \kappa + \text{interaction terms},
\ \pat = k\frac{d}{dk}.
\end{equation}
If the interaction terms are absent, the Gau\ss ian fixed point $\kappa=0$ is
the only conformal point, corresponding to a free massless theory. If the
interaction terms are nonzero, e.g., if the couplings approach interacting
fixed points by themselves, the sign of these terms decides about a possible
conformal behavior. A positive contribution from the interaction terms gives
rise to a fixed point at $\kappa>0$ which can control the conformal running
over many scales. If they are negative, no physically acceptable conformal vev
is possible. Since fermions and bosons contribute with opposite signs to the
interaction terms, the existence of a fixed point $\kappa_\ast>0$ crucially
depends on the relative strength between bosonic and fermionic fluctuations.
More specifically, the bosons have to win out over the fermions.

\begin{figure}[ht]
\centering
\includegraphics[width=0.40\textwidth]{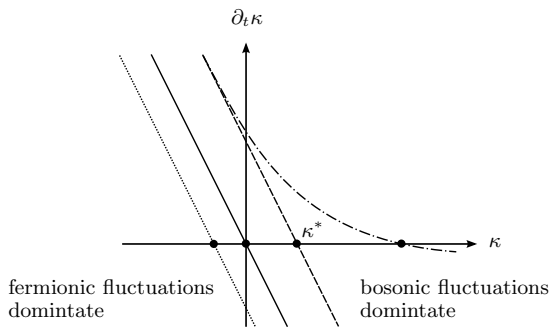}
\caption{Sketch of the flow of the dimensionless squared Higgs vacuum
  expectation value $\kappa$ as described in the text below.
  %Solid line: free massless theory, $\pat\kappa=-2\kappa$, with a trivial Gau\ss ian fixed point at $\kappa=0$. Dotted line: fermion fluctuations dominate, inhibiting an acceptable non-Gau\ss ian fixed point for positive $\kappa$.  Dashed line: bosonic fluctuations dominate, inducing a non-Gau\ss ian fixed point $\kappa^\ast>0$ with conformal behavior of the Higgs expectation value. Dot-dashed line: if the interaction terms exhibit a $\kappa$ dependence such that the slope at the fixed point (critical exponent) is reduced, the model even has an improved hierarchy behavior.
  }
\label{fig:sketch}
\end{figure}

In Fig.~\ref{fig:sketch} (taken from \cite{GiesScherer:2009}), we sketch various
options for the flow of the dimensionless squared vev $\kappa$. The solid line
depicts the free massless theory with a trivial Gau\ss ian fixed point at
$\kappa=0$ . If the fermions dominate, the interaction terms are negative and
the fixed point is shifted to negative values (being irrelevant for physics),
cf. dotted line. If the bosonic fluctuations dominate, the $\kappa$ flow
develops a non-Gau\ss ian fixed point at positive values $\kappa^\ast>0$. This
can support a conformal behavior over many orders of magnitude, cf. dashed
line. This fixed point is UV attractive, implying that the vev is a relevant
operator near the fixed point. If the interaction terms are approximately
$\kappa$ independent, the slope of $\pat\kappa$ near the fixed point is still
close to $-2$, corresponding to a critical exponent $\Theta\simeq 2$ and a
persistent hierarchy problem. An improvement of ``naturalness'' could arise
from a suitable $\kappa$ dependence of the interaction terms that results in a
flattening of the $\kappa$ flow near the fixed point, cf. dot-dashed
line. Whether or not this happens in a specific model is a prediction of the
theory which needs to be deduced from the theory by nonperturbative methods.

In this work, we construct a model with standard-model-like symmetries along
this line of research. We introduce $N_\mathrm{L}$ left-handed fermion species
$\psi^a_{\mathrm{L}}$ ($a \in \{1,...,N_\mathrm{L}\}$) and one right-handed
fermion $\psi_{\mathrm{R}}$, as well as $N_\mathrm{L}$ complex bosons
$\phi^a$. All fields live in the fundamental representation of the left-handed
chiral symmetry group U$(\NL)_\text{L}$. The Yukawa coupling is then realized
by a term $\bar{h}
(\bar{\psi}_{\mathrm{R}}\phi^{a\dagger}\psi_{\mathrm{L}}^a-\bar{\psi}_{\mathrm{L}}^a\phi^{a}\psi_{\mathrm{R}})$. This
chiral Yukawa system mimics the coupling between the standard-model Higgs
scalar and the left- and right-handed components of the top quark, also
involving Yukawa couplings to the left-handed bottom (for $\NL=2$) and further
bottom-like quarks (for $\NL>2$) in the same family. If the scalar field
develops a vev upon symmetry breaking, the top quark acquires a Dirac mass,
whereas the bottom-type quarks remain massless (similar to neutrinos in the
standard model). For $\NL=2$, we ignore the Yukawa coupling $\sim
\bar{\psi}_{\mathrm{R}} \epsilon_{ab} \phi^{a\dagger} \psi_{\mathrm{L}}^b$ in
our model, which provides for a mass term for the bottom quark in the standard
model, since it does not generalize to other $\NL$.

\begin{figure}[ht]
\centering
\includegraphics[width=0.40\textwidth]{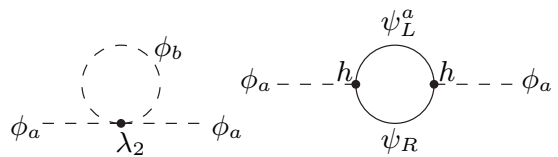}
\caption{Loop contributions to the renormalization flow of the vev. The left
  loop involves only inner boson lines. The vertex $\lambda_2$ allows for a
  coupling between all available boson components. This implies a linear
  dependence on $N_\mathrm{L}$ for the renormalization of the boson
  contribution, see below. On the right panel, we depict the corresponding
  fermion loop contribution. The incoming boson $\phi_a$ fully determines the
  structure of the fermion loop and does not allow for other left-handed inner
  fermions than $\psi_{\mathrm{L}}^a$, inhibiting an $\NL$ dependence of the algebraic
  weight of this loop.}
\label{fig:loops}
\end{figure}

In comparison to left-right symmetric models, this model has an
interesting new feature concerning the relative weight of the boson
interaction terms contributing to the renormalization of the dimensionless vev
$\kappa$ (see figure \ref{fig:loops}): diagrammatically speaking, the inner
structure of fermion/boson components in the fermion loop for a specific
choice of external boson legs is fully determined. On the other
hand, the boson loop obtains contributions from all $N_\mathrm{L}$ components
and so is linear in $N_\mathrm{L}$. In this way, $\NL$ serves as a control
parameter for boson dominance and for the potential existence of a non-Gau\ss
ian fixed point. Already at this qualitative level of the discussion, it is
worthwhile to stress that the standard model has such a left-right asymmetric
structure which can support the conformal-vev fixed point.

\subsection{Flow equation and asymptotic safety}

In the asymptotic-safety scenario of a quantum field theory, the microscopic
action to be quantized is a priori unknown. A construction of a renormalizable
theory is possible if a suitable fixed point exists in the space of all
possible action functionals, i.e., in {\em theory space}. This theory space is
spanned by all possible operators which can be constructed from the chosen
degrees of freedom and which are compatible with the desired symmetries. If
such a fixed point exists the microscopic action to be quantized can be
constructed from the properties of the fixed point and thus is a prediction
itself.
%If several fixed points exist each can serve to define a different UV completion and thus different theories. 

The framework for a nonperturbative construction of renormalizable field theories is provided by the functional RG which can be
formulated in terms of a flow equation for the effective average action
$\Gamma_k$, the Wetterich equation \cite{Wetterich:1993yh}:
\begin{equation}\label{flowequation}
	\partial_t\Gamma_k[\Phi]
        =\frac{1}{2}\mathrm{STr}\{[\Gamma^{(2)}_k[\Phi]+R_k]^{-1}(\partial_tR_k)\}.
\end{equation}
Here, $\Gamma^{(2)}_k[\Phi]$ is the second functional derivative with respect to the
field $\Phi$, the latter representing a collective field variable for all
bosonic or fermionic degrees of freedom, and $R_k$ denotes a
momentum-dependent regulator function that suppresses IR modes below a
momentum scale $k$. The solution of the Wetterich equation provides an RG
trajectory in theory space, interpolating between the bare action $S_\Lambda$ to be
quantized $\Gamma_{k\to\Lambda}\to S_\Lambda$ and the full quantum effective
action $\Gamma=\Gamma_{k\to 0}$, being the generating functional of 1PI
correlation functions; for reviews, see \cite{ReviewRG}. 

The effective average action $\Gamma_k$ can be parameterized by a possibly infinite set of generalized dimensionless couplings $g_i$. Then, the Wetterich equation
provides us with the flow of these couplings $\pat g_i=
\beta_{g_i}(g_1,g_2,\dots) $. A fixed point $g_i^\ast$ is defined by
\begin{equation}
 \beta_i(g_1^{\ast},g_2^{\ast},...)=0\ \forall \ i\,.
\end{equation}
The fixed point is non-Gau\ss ian, if at least one fixed-point coupling is
nonzero $g_j^\ast\neq 0$. If the RG trajectory flows into a fixed point in the UV, the UV cutoff can safely be taken to infinity and the theory can be considered as fundamental.

In addition to being fundamental, we also want the theory to be
predictive. For this, let us consider the fixed-point regime, where the flow
can be linearized,
\begin{equation}
 \partial_t g_i = B_i{}^j (g^\ast_j-g_j)+\dots, \quad B_i{}^j =\frac{\partial
   \beta_{g_i}}{\partial g_j} \Big|_{g=g^\ast}.\label{eq:lin}
\end{equation}
The critical exponents $\Theta^I$ correspond to the negative of the
eigenvalues of the stability matrix $B_i{}^j$. They allow for a classification of
physical parameters: Whereas all eigendirections with $\Theta^I<0$ die out towards
the IR and thus are irrelevant, all relevant directions with $\Theta^I>0$
increase towards the IR and thus determine the macroscopic physics (for the
marginal directions $\Theta^I=0$, it depends on the higher-order terms in the
expansion about the fixed point). Hence the number of relevant and
marginally-relevant directions determines the number of physical parameters to
be fixed. The theory is predictive if this number is finite. In the case of
the Gau\ss ian fixed point $g_i^\ast=0$, the present construction corresponds
to the standard perturbative power-counting analysis and the critical
exponents are equal to the power counting dimensions of the couplings.

If a critical exponent is much larger than zero, say of $\mathcal O(1)$, the
RG trajectory rapidly leaves the fixed-point regime towards the IR. Therefore,
separating a typical UV scale where the system is close to the fixed point
from the IR scales where, e.g., physical masses are generated requires a
significant fine-tuning of the initial conditions.

For the flow towards the IR, the linearized fixed-point flow \Eqref{eq:lin}
generally is insufficient and the full nonlinear $\beta$ functions have to be
taken into account. Even the parameterization of the effective action in terms
of the same degrees of freedom in the UV and IR might be
inappropriate. Nevertheless, we use the same bosonic and fermionic degrees of
freedom on all scales in the present work, since we specifically want to
address the question whether standard-model IR degrees of freedom can have an
interacting UV completion.

\section{Renormalization flow of chiral Yukawa systems}
\label{sec:RG}

\subsection{Derivative expansion}\label{general:flowequations}

In the present work, we investigate a Yukawa theory with chiral fermions
including one right-handed fermion and $N_\mathrm{L}$ left-handed
fermions. The fermions are coupled to $N_\mathrm{L}$ complex bosons via a
simple Yukawa interaction. We span the theory space by a truncated action
functional in a derivative expansion, which reads at next-to-leading order
\begin{eqnarray}\label{eq:SULtruncation}
\Gamma_k&=&\int d^dx\Big\{i(Z_{\mathrm{L},k}\bar{\psi}_{\mathrm{L}}^a\dslash\psi_{\mathrm{L}}^a
+Z_{\mathrm{R},k}\bar{\psi}_{\mathrm{R}}\dslash\psi_{\mathrm{R}})\nonumber\\ 
&{}&+Z_{\phi , k}(\partial_{\mu}\phi^{a\dagger})(\partial^{\mu}\phi^a)
+U_k(\phi^{a\dagger}\phi^a)\nonumber\\
&{}&+\bar h_k\bar{\psi}_{\mathrm{R}}\phi^{a\dagger}\psi_{\mathrm{L}}^a-\bar
h_k\bar{\psi}_{\mathrm{L}}^a\phi^{a}\psi_{\mathrm{R}}\Big\}. 
\end{eqnarray}
The fermion fields $\psi_{\mathrm{L}}^a$ and $\psi_{\mathrm{R}}$ have standard kinetic terms but can
acquire different wave function renormalizations $Z_{\mathrm{L},k}$ and $Z_{\mathrm{R},k}$. The
index $a$ runs from $1$ to $N_\mathrm{L}$. The projections on the
left-/right-handed fermion contributions are carried out via the projection
operators
\begin{equation}
P_{\mathrm{L/R}}=\frac{1}{2}(1\pm \gamma_5)\,.
\end{equation}
The bosonic sector involves a standard kinetic term with wave function
renormalization $Z_{\phi , k}$ and an effective potential
$U_k(\phi^{a\dagger}\phi^a)$. Defining the invariant
$\rho:=\phi^{a\dagger}\phi^a$, the effective potential $U_k(\rho)$ can be
expanded in powers of $\rho$. The bosons can also be expressed in terms of a
real field basis by defining
\begin{eqnarray}
 \phi^a = \frac{1}{\sqrt{2}}(\phi_1^a + i\phi_2^a),\quad \phi^{a\dagger} = \frac{1}{\sqrt{2}}(\phi_1^a - i\phi_2^a)\,,
\end{eqnarray}
where $\phi_1^a, \phi_2^a\in\mathbbm{R}$. The truncated effective action
\eqref{eq:SULtruncation} including the  Yukawa interaction is invariant under
U$(N_\mathrm{L})_\mathrm{L}$ transformations of the left-handed fermion and
the boson as well as U$(1)_\mathrm{R}$ transformations of the right-handed
fermion and the boson. 

All the parameters in the effective average action are understood to be scale
dependent, which is indicated by the index $k$.

The flow of the wave function renormalizations $Z_{\phi , k},
Z_{\mathrm{L},k}$ and $Z_{\mathrm{R},k}$ can be expressed in terms of
scale-dependent anomalous dimensions
\begin{equation}
	\eta_{\phi}=-\partial_t \mbox{ln} Z_{\phi,k},\ 
        \eta_{\mathrm{L,R}}=-\partial_t \mbox{ln} Z_{\mathrm{L,R},k}\,.
\end{equation}
Setting the anomalous dimensions to zero defines the leading-order derivative
expansion. At next-to-leading order, it is important to distinguish
between $Z_{\mathrm{L},k}$ and $Z_{\mathrm{R},k}$ as they acquire different loop
contributions, see below.

In order to fix the standard RG invariance of field rescalings, we define the
renormalized fields as
\begin{equation}
 	\tilde{\phi}=Z_{\phi,k}^{1/2}\phi,\ \tilde{\psi}_{\mathrm{L,R}}=Z_{\mathrm{L,R}}^{1/2}\psi_{\mathrm{L,R}}.
\end{equation}
For the fixed-point search, it is useful to introduce dimensionless
renormalized quantities
\begin{eqnarray}\label{eq:dimensionless}
\tilde{\rho}&=&Z_{\phi,k}k^{2-d}\rho,\\
 h_k^2&=&Z_{\phi,k}^{-1}Z_{\mathrm{L},k}^{-1}Z_{\mathrm{R},k}^{-1}k^{d-4}\bar h_k^2 ,\\
 u_k(\tilde\rho)&=&k^{-d}U_k(\rho)|_{\rho=k^{d-2}\tilde\rho/Z_{\phi,k}}.
\end{eqnarray}
Detailed information about the derivation of the flow equations for this
truncation in arbitrary spacetime dimensions $d$ is given in
App.~\ref{sec:deriv-effect-potent}. For our purposes, we use a linear
regulator function $R_k$ which is optimized for the present truncation
\cite{Litim:2001up}. The flow of the effective potential in terms of threshold
functions which are given in App.~\ref{section:threshold} reads
\begin{eqnarray}\label{basic:flowequation}
\partial_t u_k&=&-d u_k+\tilde{\rho}u_k'(d-2+\eta_{\phi})\\
&+&2v_d\big\{(2N_{\mathrm{L}}-1)l_0^d(u_k')+l_0^d(u_k'+2 \tilde{\rho}u_k'')\nonumber\\
&-&d_{\gamma}\big((N_{\mathrm{L}}-1)l_{0,\mathrm{L}}^{(\mathrm{F})d}(0)+l_{0,\mathrm{L}}^{(\mathrm{F})d}(\tilde{\rho}h_k^2)+l_{0,\mathrm{R}}^{(\mathrm{F})d}(\tilde{\rho}h_k^2)\big) \big\},\nonumber
\end{eqnarray}
where the primes denote derivatives with respect to $\tilde\rho$, and
$v_d=1/(2^{d+1}\pi^{d/2} \Gamma(d/2))$. For the symmetric phase (SYM), we
expand the effective potential around zero field,
\begin{eqnarray}\label{eq:symeffpot}
  u_k&=&\sum_{n=1}^{N_p}\frac{\lambda_{n,k}}{n!}\tilde{\rho}^n 
  = m_k^2\tilde{\rho}+\frac{\lambda_{2,k}}{2!}\tilde{\rho}^2
  +\frac{\lambda_{3,k}}{3!}\tilde{\rho}^3+.... 
\end{eqnarray}
For the SSB phase, where the minimum of the effective potential $u_k$ acquires
a nonzero value $\kappa_k:=\tilde\rho_{\mbox{min}}> 0$, we use the expansion
\begin{eqnarray}
 	u_k&=&\sum_{n=2}^{N_\text{p}}\frac{\lambda_{n,k}}{n!}(\tilde{\rho}-\kappa_k)^n\label{eq:uexpSSB}\\ &=&\!\frac{\lambda_{2,k}}{2!}(\tilde{\rho}-\kappa_k)^2
        \!+\frac{\lambda_{3,k}}{3!}(\tilde{\rho}-\kappa_k)^3+...\nonumber
\end{eqnarray}
Given the flow of $u_k$ \eqref{basic:flowequation}, the flows of $m_k^2$ or
$\lambda_{n,k}$ in both phases can be read off from an expansion of the flow
equation and a comparison of coefficients. For the flow of $\kappa_k$, we use
the fact that the first derivative of $u_k$ vanishes at the minimum,
$u_k'(\kappa_k)=0$. This implies
\begin{eqnarray}
 0= \pat u_k'(\kappa_k)&=&\partial_t u_k'(\tilde\rho)|_{\tilde\rho=\kappa_k}
  +(\partial_t \kappa_k)u_k''(\kappa_k)\nonumber\\
	\Rightarrow \partial_t \kappa_k&=&-\frac{1}{u_k''(\kappa_k)}\partial_t
        u_k'(\tilde\rho)|_{\tilde\rho=\kappa_k}\,. \label{eq:kappa}
\end{eqnarray}
The explicit flow equations for the running parameters will be given in the
following sections for the SYM and the SSB phase separately. Note that the expansion coefficients $\lambda_{n,k}$ in Eqs. \eqref{eq:symeffpot} and \eqref{eq:uexpSSB} are not identical. Since there is little risk that the notation of the different regimes interferes with each other, we refrain from introducing different symbols.

In the SSB regime, the flow of the Yukawa coupling and the scalar anomalous
dimension for the Goldstone mode can, in principle, be different from that of
the radial mode. As the Goldstone modes as such are not present in the
standard model, we compute the Yukawa coupling and the scalar anomalous
dimension by projecting the flow onto the radial scalar operators in the SSB
regime. Note that this strategy is different from that used for critical
phenomena in other Yukawa or bosonic systems, where the Goldstone modes can
dominate criticality. 

Accordingly, the flow of the Yukawa coupling $h_k$ can be derived (see
App.~\ref{section:yukawaflow}), and we end up with
\begin{widetext}
\begin{eqnarray}
\partial_t h_k^2 &=&(d-4+\eta_{\phi}+\eta_{\mathrm{L}}+\eta_{\mathrm{R}})h_k^2+ 4v_d h_k^4\Big\{
(2\tilde{\rho}u_{k}'') l_{1,2}^{(\mathrm{FB})d}(\tilde{\rho}h_k^2, u_{k}')-(6\tilde{\rho}u_{k}''+4\tilde{\rho}^2u_{k}''')l_{1,2}^{(\mathrm{FB})d}(\tilde{\rho}h_k^2, u_{k}'+2\tilde{\rho}u_{k}'')\label{eq:flowh2}\\
&{}&-l_{1,1}^{(\mathrm{FB})d}(\tilde{\rho}h_k^2, u_{k}')+l_{1,1}^{(\mathrm{FB})d}(\tilde{\rho}h_k^2, u_{k}'+2\tilde{\rho}u_{k}'')+2\tilde{\rho}h_k^2l_{2,1}^{(\mathrm{FB})d}(\tilde{\rho}h_k^2, u_{k}')-2\tilde{\rho}h_k^2l_{2,1}^{(\mathrm{FB})d}(\tilde{\rho}h_k^2, u_{k}'+2\tilde{\rho}u_{k}'')\Big\}.\nonumber
\end{eqnarray}
Finally, we list the expressions for the anomalous dimensions
\begin{eqnarray}
\eta_{\phi}&=&\frac{8 v_d}{d}\tilde\rho(3 u_k'' +2\tilde\rho
u_k''')^2m_{22}^d(u_k'+2\tilde\rho u_k'')+\label{eq:etaphi}\frac{(2N_{\mathrm{L}}-1)8v_d}{d}\tilde\rho u_k''^2 m_{22}^d(u_k')\\
&{}&+\frac{8v_d d_\gamma}{d}h_k^2 m_4^{(\mathrm{F})4}(\tilde\rho h_k^2)+\frac{8v_d d_\gamma}{d}\tilde\rho h_k^4m_2^{(\mathrm{F})4}(\tilde\rho h_k^2),\nonumber\\
\eta_{\mathrm{L}}&=&\frac{8v_d}{d}h_k^2[ m_{12}^{(\mathrm{FB})d}(\tilde{\rho}h_k^2,
  u_{k}'+2\tilde{\rho} u_{k}'')\label{eq:etaL}+m_{12}^{(\mathrm{FB})d}(\tilde{\rho}h_k^2, u_{k}') ],\\
\eta_{\mathrm{R}}&=&\frac{8v_d}{d}h_k^2[ m_{12}^{(\mathrm{FB})d}(\tilde{\rho}h_k^2,
  u_{k}'+2\tilde{\rho} u_{k}'')\label{eq:etaR}+m_{12}^{(\mathrm{FB})d}(\tilde{\rho}h_k^2, u_{k}') +2(N_{\mathrm{L}}-1)m_{12}^{(\mathrm{FB})d}(0, u_{k}')].
\end{eqnarray}
\end{widetext}
The arguments of the threshold functions have to be evaluated at the
minimum of the effective potential. In the following sections, we will
concentrate on the case of $d=4$ dimensions. 

Let us emphasize that the derivative expansion has already been tested in
various Yukawa systems and has proved to yield qualitatively and
quantitatively accurate results. RG flows for Yukawa systems have been successfully
studied in QCD \cite{Jungnickel:1995fp}, critical phenomena
\cite{Rosa:2000ju}, quantum phase transitions \cite{Strack} and ultra-cold
fermionic atom gases \cite{Birse:2004ha}.

\subsection{Parameter constraints}

Let us finally discuss several constraints on the couplings as, e.g., dictated
by physical requirements as well as by our truncation. As our truncation is
based on a derivative expansion, satisfactory convergence is expected if the
higher derivative operators take little influence on the flow of the
leading-order terms. In the present case, the leading-order effective
potential receives higher-order contributions only through
the anomalous dimensions. Therefore, convergence of the derivative expansion
requires
\begin{equation}
 \eta_{\text{L}}, \eta_{\text{R}}, \eta_\phi \lesssim \mathcal O(1).\label{eq:validity}
\end{equation}
This condition will serve as an important quality criterion for our
truncation.  The SYM regime is characterized by a minimum of $u_k$ at
vanishing field. A simple consequence is that the mass term needs to be
positive. Also, the potential should be bounded from below, which in the
polynomial expansion translates into a positive highest nonvanishing
coefficient,
\begin{equation}
m^2_k,\,  \lambda_{n_{\text{max}},k}>0. \label{eq:constraint1}
\end{equation}
In the SSB regime, the minimum must be positive, $\kappa_k>0$, the potential
should be bounded, and in addition the potential at the minimum must have
positive curvature,
\begin{equation}
\kappa_k, \, \lambda_{n_{\text{max}},k},\, \lambda_{2,k}>0.\label{eq:constraint2}
\end{equation}
Finally, Osterwalder-Schrader positivity requires 
\begin{equation}
h_k^2>0.\label{eq:hcrit}
\end{equation}
Beyond that, there are no constraints on the size of the couplings as in
perturbation theory.

\section{The Symmetric Regime (SYM)}
\label{sec:SYM}

Let us first investigate the fixed-point structure of the system in the
symmetric regime. We restrict ourselves to four dimensions and expand the
effective action using the ansatz \eqref{eq:symeffpot}. Now and in the
following we will suppress the index $k$ at the parameters of the truncation
for notational convenience. The flow equations are evaluated at the minimum of
the effective potential, which is at $\tilde\rho=0$ in the symmetric regime,
and we replace $u'=m^2$, $u''=\lambda_2$,
$u'''=\lambda_3, ...\,$. In this case, the next-to-leading
order flow equations up to second order in the effective potential read
\begin{eqnarray}
\partial_tm^2&=&(\eta_{\phi}-2)m^2-\frac{1}{16\pi^2}(1-\frac{\eta_{\phi}}{6})\frac{\lambda_2(N_{\mathrm{L}}+1)}{(1+m^2)^2}\\
&&+\frac{h^2}{8\pi^2}(1-\frac{\eta_{\mathrm{L}}}{5})+\frac{h^2}{8\pi^2}(1-\frac{\eta_{\mathrm{R}}}{5}),\nonumber\\
\partial_t\lambda_2&=&2\eta_{\phi}\lambda_2+\frac{1}{8\pi^2}(1-\frac{\eta_{\phi}}{6})\frac{\lambda_2^2(N_{\mathrm{L}}+4)}{(1+m^2)^3}\\
&&-\frac{1}{16\pi^2}(1-\frac{\eta_{\phi}}{6})\frac{\lambda_3(N_{\mathrm{L}}+2)}{(1+m^2)^2}\nonumber\\
&&-\frac{h^4}{4\pi^2}(1-\frac{\eta_{\mathrm{L}}}{5})-\frac{h^4}{4\pi^2}(1-\frac{\eta_{\mathrm{R}}}{5})\,.\nonumber
\end{eqnarray}
We observe that the flow equation of a coupling $\lambda_n$ involves $m^2,
\lambda_2,...,\lambda_{n+1}$.  For the Yukawa coupling we obtain
\begin{equation}
\partial_t h^2=(\eta_{\phi}+\eta_{\mathrm{L}}+\eta_{\mathrm{R}})h^2\,.
\end{equation}
In particular, all contributions from typical vertex triangle diagrams vanish
in the SYM regime. At leading
order in a derivative expansion where $\eta_i=0$, the Yukawa coupling does not
flow in the SYM regime. At next-to-leading order, we also have to take into
account the anomalous dimensions
\begin{eqnarray}\label{eq:anomdimsym}
\eta_{\phi}&=&\frac{1}{4\pi^2}h^2 \left( 1-\frac{\eta_{\mathrm{L}}+\eta_{\mathrm{R}}}{8}\right),\\
\eta_{\mathrm{L}}&=&\frac{1}{16\pi^2}h^2\left(1-\frac{\eta_{\phi}}{5}\right)\frac{2}{(1+m^2)^2},\\
\eta_{\mathrm{R}}&=&\frac{1}{16\pi^2}h^2\left(1-\frac{\eta_{\phi}}{5}\right)\frac{2N_{\mathrm{L}}}{(1+m^2)^2}\,.
\end{eqnarray}
For an interacting fixed point with a non-vanishing $h^2$ to exist, the
prefactor $(\eta_{\phi}+\eta_{\mathrm{L}}+\eta_{\mathrm{R}})$ in the flow equation for the
Yukawa coupling would have to vanish at some point in parameter space. This
implies that the different anomalous dimensions need to have a relative
minus sign. 
From \Eqref{eq:anomdimsym}, we read off that a relative sign change requires
either $\eta_\phi>5$ or $\eta_{\mathrm{L}}+\eta_{\mathrm{R}}>8$. Both options would clearly be in
conflict with the validity bounds of the derivative expansion, demanding at
least for $\eta_{\mathrm{L,R},\phi} \lesssim \mathcal O(1)$. 

Therefore, a nontrivial fixed point of the Yukawa coupling in the SYM
regime based on the criterion $\eta_{\phi}+\eta_{\mathrm{L}}+\eta_{\mathrm{R}}=0$ -- even if it
existed -- would be beyond the reliability bounds of the derivative
expansion. Within the validity regime of our truncation, we thus find only the
Gau\ss ian fixed point for the Yukawa coupling $h^\ast=0$. Since the flow of
the Yukawa coupling is proportional to itself, an initial zero value for $h^2$
will leave the system non-interacting at $h^2=0$ for all scales. The remaining
interacting system is purely bosonic and does not show any nontrivial fixed
point in agreement with triviality of $\phi^4$ theory. We conclude that the
chiral Yukawa model in the SYM regime is not asymptotically safe within the
part of theory space spanned by our truncation. 

% old removed parts can be found below \end{document}

\section{The Regime of Spontaneous Symmetry Breaking (SSB)}
\label{sec:SSB}

Let us continue our fixed-point search in the broken-symmetry regime. The flow
equations have a richer structure here, since the coupling to the condensate
can mediate further effective interactions. Most importantly, the broken
regime can support the fixed-point scenario with a conformal vev. 

\subsection{Fixed-point search to leading order}\label{basic} 

For the fixed-point search, we use a polynomial expansion of the effective
potential about its minimum, cf. \Eqref{eq:uexpSSB}. As a check of the quality
of this expansion, the convergence properties of physical quantities have to be
determined with respect to increasing orders in this expansion. In general, we
expect that this expansion gives a good approximation of the full effective
potential only in the vicinity of the potential minimum. Nevertheless, this can
still lead to sufficient information for a quantitative estimate of a variety
of physical quantities, as they are mostly related to properties of the
potential near the minimum. 

For a first glance, we use the simplest nontrivial approximation of the
effective potential, $u=\frac{\lambda_2}{2}(\tilde{\rho}-\kappa)^2$, which
facilitates a purely analytical treatment of the fixed-point equations.  The
flow of the effective potential \eqref{basic:flowequation}, evaluated around the minimum, then boils down to
flows for $\tilde\rho_\mathrm{min}=\kappa$ and
$u''=\lambda_2$, with all other couplings being zero, i.e.,
$u'=0$, and $u^{(n)}=0$ for $n>2$.  We also
confine ourselves to the leading-order approximation in the derivative
expansion, where the anomalous dimensions vanish,
$\eta_{\text{L,R},\phi}=0$. The flow equations in this approximation read
explicitely
\begin{widetext}
\begin{eqnarray}
\partial_t\kappa&=&-2\kappa+\frac{1}{32\pi^2}(2N_{\mathrm{L}}-1)
+\frac{3}{32\pi^2}\frac{1}{(1+2\kappa\lambda_2)^2}-\frac{1}{4\pi^2}\frac{h^2}{\lambda_2(1+\kappa  h^2)^2},\\ 
\partial_t\lambda_2&=&\frac{1}{16\pi^2}(2N_{\mathrm{L}}-1)\lambda_2^2
+\frac{1}{16\pi^2}\frac{9\lambda_2^2}{(1+2\kappa\lambda_2)^3}-\frac{1}{2\pi^2}\frac{h^4}{(1+\kappa h^2)^3},
\end{eqnarray}
and for the Yukawa coupling
\begin{eqnarray}
\partial_t h^2&=&\frac{1}{16\pi^2}\frac{h^4}{(1+\kappa h^2)}\Bigg\{2\lambda_2\kappa\left( \frac{1}{1+\kappa h^2} +2\right)-\frac{6\kappa\lambda_2}{(1+2\kappa\lambda_2)^2}\left( \frac{1}{1+\kappa h^2} + \frac{2}{1+2\kappa\lambda_2}\right)\\
&{}&-\left( \frac{1}{1+\kappa h^2}+ 1\right)+\frac{1}{(1+2\kappa\lambda_2)}\left( \frac{1}{1+\kappa h^2} + \frac{1}{1+2\kappa\lambda_2} \right)\nonumber\\
&{}&+\frac{2\kappa h^2}{(1+\kappa h^2)}\left( \frac{2}{1+\kappa h^2}+1 \right)-\frac{2\kappa h^2}{(1+\kappa h^2)(1+2\kappa\lambda_2)}\left( \frac{2}{1+\kappa h^2}+\frac{1}{1+2\kappa\lambda_2}\right) \Bigg\}\nonumber\,.
\end{eqnarray}

\begin{figure}
\begin{center}
 \includegraphics[width=0.3\textwidth]{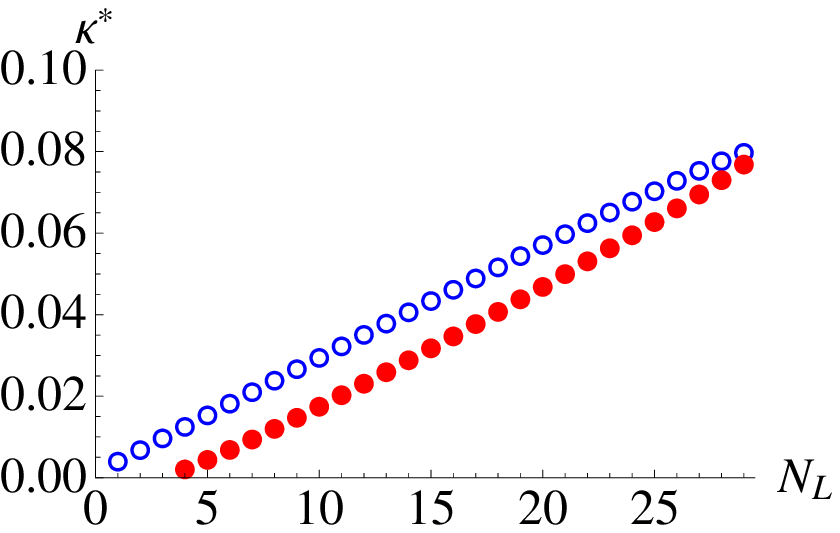}
 \includegraphics[width=0.3\textwidth]{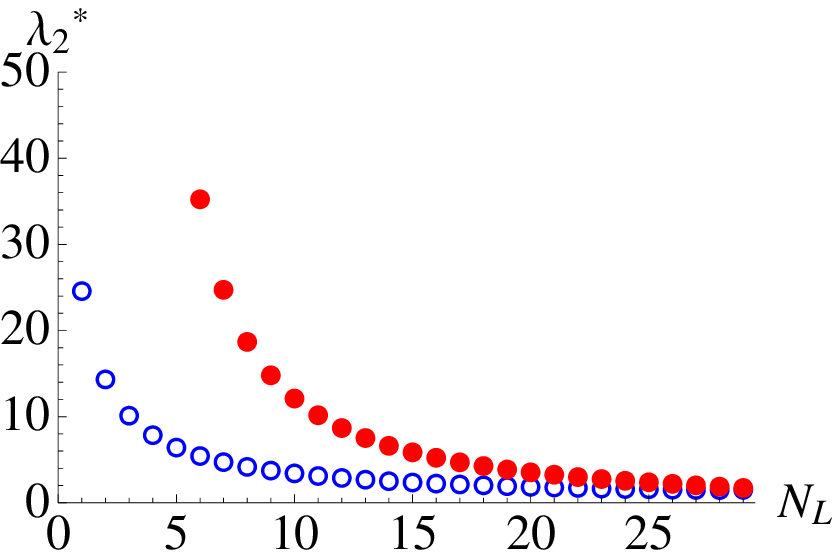}
 \includegraphics[width=0.3\textwidth]{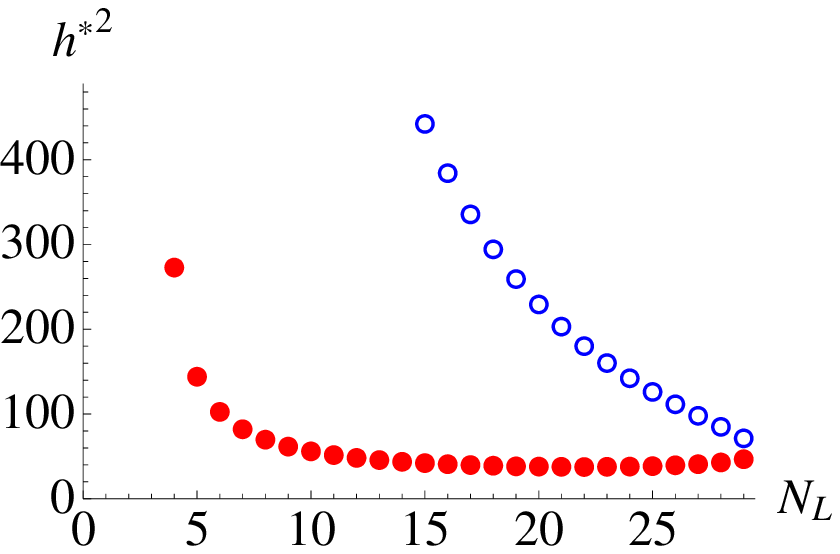}
 \caption{Leading-order derivative expansion (local-potential approximation)
   as well as leading-order polynomial expansion: fixed-point values
   $\{\kappa^*,\lambda_2^*,h^{*2}\}$ for the two physically admissible fixed
   points as a function of $N_\mathrm{L}$. } 
\label{fig:ssbfix}
\end{center}
\end{figure}

\begin{figure}
\begin{center}
 \includegraphics[width=0.3\textwidth]{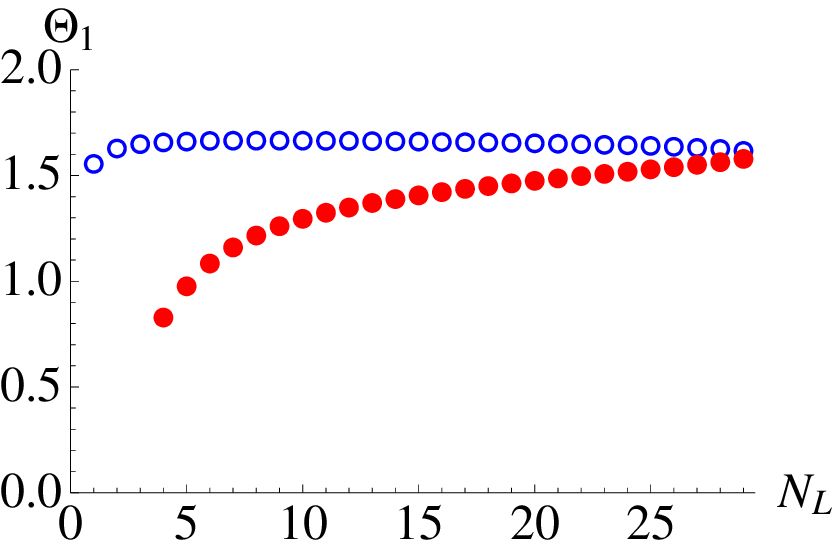}
 \includegraphics[width=0.3\textwidth]{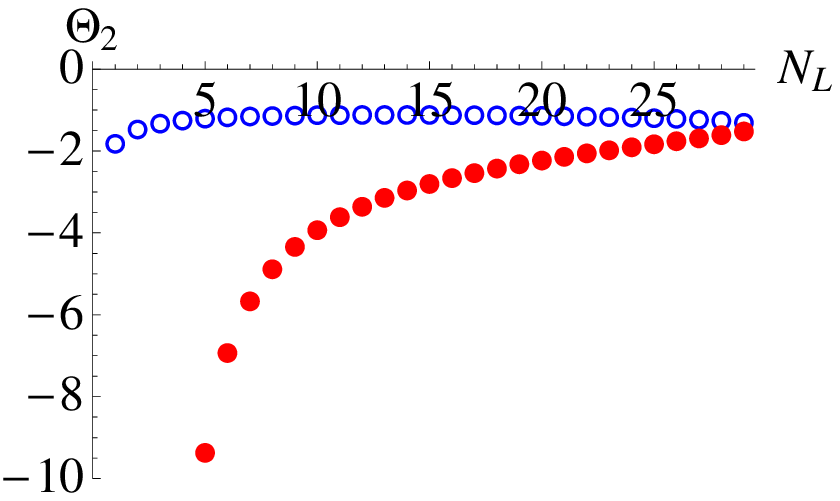}
 \includegraphics[width=0.3\textwidth]{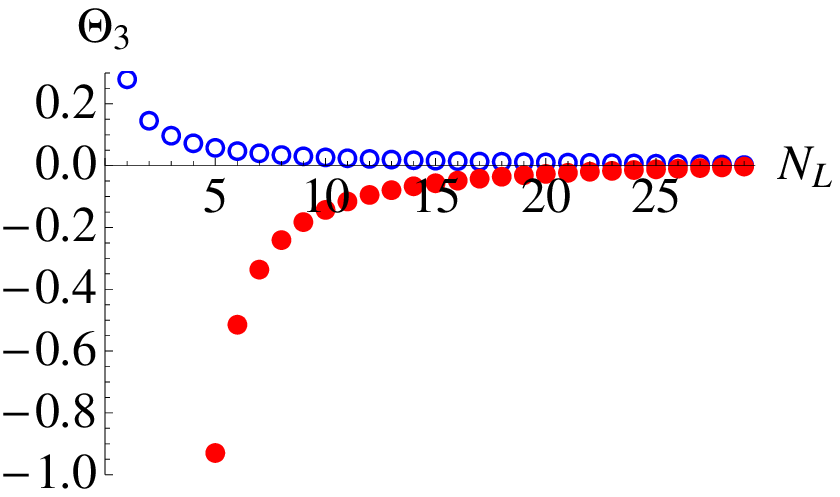}
 \caption{Leading order truncation: Critical exponents as a function of
   $N_\mathrm{L}$. The fixed point corresponding to the open circles has two relevant directions, whereas the fixed point corresponding to the filled circles has only one relevant direction.}
\label{fig:ssbtheta}
\end{center}
\end{figure}

\end{widetext}
The resulting set of fixed-point equations
$\{\partial_t\kappa=0,\partial_t\lambda_2=0,\partial_t h^2=0\}$ can be solved
analytically. For a given $N_\mathrm{L}$, we obtain a large number of fixed
points but only up to two fixed points fulfill the physical constraints given
in section \ref{general:flowequations}. For $N_\mathrm{L} < 4$, we find only
one admissible non-Gau\ss ian fixed point (NGFP). For $4 \leq N_\mathrm{L}
\leq 29$, two admissible NGFPs occur. For $30 \leq N_\mathrm{L} \leq 57$,
there is again only one NGFP, whereas larger $N_\mathrm{L}$ do not give rise
to any physically admissible fixed points. The fixed-point values
$\kappa^\ast$, $h^\ast{}^2$, and $\lambda_2^\ast$ as a function of
$N_\mathrm{L}$ can be read off from figure \ref{fig:ssbfix}.

The universal critical exponents can be deduced from the linearized flow
around this fixed point, cf. \Eqref{eq:lin},
\begin{equation}
 \partial_t g_i = B_i{}^j (g^\ast_j-g_j)+\dots, \label{eq:lin2}
\end{equation}
where $g_i=(\kappa,\lambda_2,h^2)$ in the present simple truncation. The
expansion coefficients ${B}_i{}^j$ form the stability matrix, the eigenvalues
$-\Theta_i$ of which correspond to the critical exponents apart from a minus
sign. The critical exponents as a function of $N_\mathrm{L}$ are shown in
Fig.~\ref{fig:ssbtheta} for this simple truncation.
It is remarkable that one NGFP has two relevant directions and the other even
only one relevant direction with a positive critical exponent. This implies
that the corresponding model is fixed by only two or one physical parameter,
respectively. This should be compared to the Gau\ss ian fixed point with
perturbative critical exponents of the order $\Theta_\kappa\simeq2$,
$\Theta_{\lambda_2}\simeq0$, and $\Theta_h\simeq 0$, implying that 3 physical
parameters need to be fixed for Gau\ss ian models. Moreover, both NGFPs
exhibit a largest critical exponent being smaller than the perturbative
maximal value, $\Theta_{\text{max}}<2$. As a consequence, these models defined
at the NGFPs have an improved hierarchy behavior.

\subsection{Benchmark fixed point for $\NL=10$}\label{basic10FP}

For further investigations and by way of example, we concentrate on one
particular fixed point. We use a fixed point with only one relevant direction,
as it is phenomenologically most appealing due to the reduction of physical
parameters. We choose $\NL=10$ in the following, since the corresponding
fixed point does not give rise to extreme coupling values, implying numerical
stability. In addition, this particular value of $\NL$ may also be of
relevance for certain unification scenarios. 

Furthermore, we extend the lowest-order polynomial expansion studied above to
higher operators, cf. \Eqref{eq:uexpSSB},
\begin{eqnarray}
 	u&=&\sum_{n=2}^{N_\text{p}} \frac{\lambda_n}{n!}
          (\tilde{\rho}-\kappa)^n\label{eq:uexpSSB2}
	  ,
\end{eqnarray}
where $N_\text{p}$ labels the higher orders. From a technical viewpoint, the
fixed-point analysis becomes much more involved at higher orders. In
particular, we have not found any analytical solutions to the higher-order
fixed-point equations. Due to the intrinsic nonlinearity of the fixed-point
equations, also a numerical search is nontrivial: a complete identification of
all possible fixed points appears out of reach.  

\begin{figure}
\begin{center}
 \includegraphics[width=0.4\textwidth]{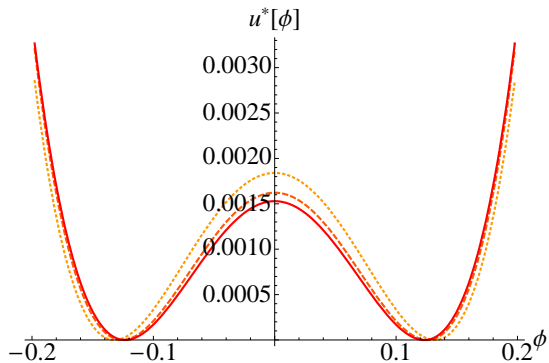}
 \caption{Convergence of the fixed-point potential $u^*$ upon the inclusion
   of higher orders in the polynomial expansion of the effective potential.
   The dotted line shows the fixed-point potential within the simple
   truncation of Subsect.~\ref{basic}, including only the couplings $\kappa$
   and $\lambda_2$. The dashed curve also includes $\lambda_3$ and $\lambda_4$,
   and the solid line extends up to $\lambda_6$.}
\label{fig:effpotplot}
\end{center}
\end{figure}

Therefore, we start with the assumption that the simple truncation involving
only $h^2, \lambda_2$ and $\kappa$ gives already a satisfactory fixed-point
estimate. As a next step, we can search for a fixed-point of the subsequent
higher-order system in the vicinity of the previous lower-order fixed
point. This procedure turns out to be self-consistent and can be iterated to
higher orders. The results for the associated fixed-point values up to
$N_{\text{p}}=6$ are summarized in Tab.~\ref{tab:fpvalues1}.
\begin{table}[ht]
\footnotesize
\centering
	\begin{tabular}{c|c|c|c|c|c|c|c}
	$N_\mathrm{L}=10$ &$h^2_\ast$&	$\kappa_\ast$	&$\lambda_2^\ast$	  & $\lambda_3^\ast$ & $\lambda_4^\ast$ & $\lambda_5^\ast$ & $\lambda_6^\ast$\\
	\cline{1-8}
	$N_\text{p}=2$ & 55.8 & 0.0174 & 12.11 & - & - & - &- \\
	$N_\text{p}=4$ & 56.0 & 0.0158 & 12.09 & -115 & $1.30\cdot10^4$ & - &- \\
	$N_\text{p}=6$ & 57.4 & 0.0152 & 12.13 & -152 & $1.20\cdot10^4$ & $-8.76\cdot10^5$ & $1.44\cdot10^8$
	\end{tabular}
\caption{Fixed-point values at different orders in the polynomial expansion of
  the effective potential. The index $N_{\text{p}}$ denotes the largest exponent of
  $\rho$ occurring in the polynomial expansion (\ref{eq:uexpSSB}) or (\ref{eq:uexpSSB2}). We observe satisfactory
  convergence of the fixed point values.}
\label{tab:fpvalues1}
\end{table}
The resulting fixed-point values show a satisfactory convergence behavior,
indicating that the polynomial expansion is quantitatively reliable already at
low orders. This convergence property is also confirmed by the form of the
fixed-point potential in a finite environment of the expansion point $\kappa$,
as is depicted in Fig.~\ref{fig:effpotplot}: The fixed-point potential of
Mexican-hat type remains quantitatively stable under the inclusion of higher
orders -- even away from the minimum $\kappa$. 

Whereas the fixed-point potential is scheme dependent, also the stability of
the universal critical exponents can be checked under the inclusion of higher
orders in the polynomial expansion. The results for the associated critical
exponents up to $N_{\text{p}}=6$ are summarized in
Tab.~\ref{tab:fpvalues2}. The leading exponents vary only on the 10\%-level
upon the inclusion of higher-order terms. Most importantly, the number of
relevant directions is not changed at higher orders, and the maximal critical
exponent tends to smaller values at higher orders which leads to an improved
hierarchy behavior. We conclude that the fixed point exists at leading-order
in the derivative expansion. It possesses many desirable properties and can
well be approximated by a polynomial expansion.

\begin{table}[ht]
\footnotesize
\centering
	\begin{tabular}{c|c|c|c|c|c|c|c}
	$N_\mathrm{L}=10$ &$\Theta_{1}$&	$\Theta_{2}$	&$\Theta_{3}$	  & $\Theta_{4}$ & $\Theta_{5}$ & $\Theta_{6}$ & $\Theta_{7}$\\
	\cline{1-8}
	$N_{\text{p}}=2$ & 1.294 & -0.143 & -3.94 & - & - & - &- \\
	$N_{\text{p}}=4$ & 1.167 & -0.170 & -2.50 & -5.53 & -13.61 & - &- \\
	$N_{\text{p}}=6$ & 1.056 & -0.175 & -2.35 & -4.97 & -8.49 & -14.02 & -25.54
	\end{tabular}
\caption{Critical exponents in the polynomial expansion of the effective
  potential. The index $N_{\text{p}}$ denotes the largest exponent of $\rho$
  occurring in the polynomial expansion (\ref{eq:uexpSSB}) or (\ref{eq:uexpSSB2}). We observe a reasonable
  convergence of the universal critical exponents.  }
\label{tab:fpvalues2}
\end{table}

\section{Predictive power of asymptotically-safe Yukawa systems}
\label{sec:PP}

We illustrate the predictive power of an asymptotically safe Higgs sector,
using the flow of our system in the local-potential approximation as a model
in its own right.

%\subsection{Higgs mass from asymptotic safety}

Perturbatively, our model would be defined by three bare parameters if the
microscopic theory is initiated near the Gau\ss ian fixed point. These three
parameters can be specified, e.g., in terms of the bare Yukawa coupling $\bar h$,
the $\phi^4$ coupling $\bar\lambda_2$, and the scalar mass term in units of the
UV cutoff $\bar m^2/\Lambda^2$. \footnote{Owing to triviality at the Gau\ss
  ian fixed point, the cutoff cannot be removed. Strictly speaking, this leads
  to a lack of universality, such that the couplings of higher-dimensional
  operators also correspond to physical parameters. Their influence on the IR
  physics is, however, suppressed by corresponding inverse powers of the UV
  cutoff $\Lambda$.}

The bare parameters can be fixed in terms of renormalization conditions. The
latter ultimately relates these parameters to the physical IR parameters: the
Higgs vev $v$, the Higgs mass $m_{\text{Higgs}}$ and the top mass
$m_{\text{top}}$. These, in turn, are related to the renormalized couplings
$\kappa,h^2, \lambda_2$ by
\begin{equation}\label{eq:topmass3}
v=\lim_{k\to0} \sqrt{2\kappa}\;k, \quad m_{\text{top}}=\sqrt{ h^2 }\; v
, \quad 
 m_{\text{Higgs}}=\sqrt{ \lambda_2 }\; v.
\end{equation}
(The bottom-type quarks remain massless in our model.)
\begin{figure}
\begin{center}
 \includegraphics[width=0.3\textwidth]{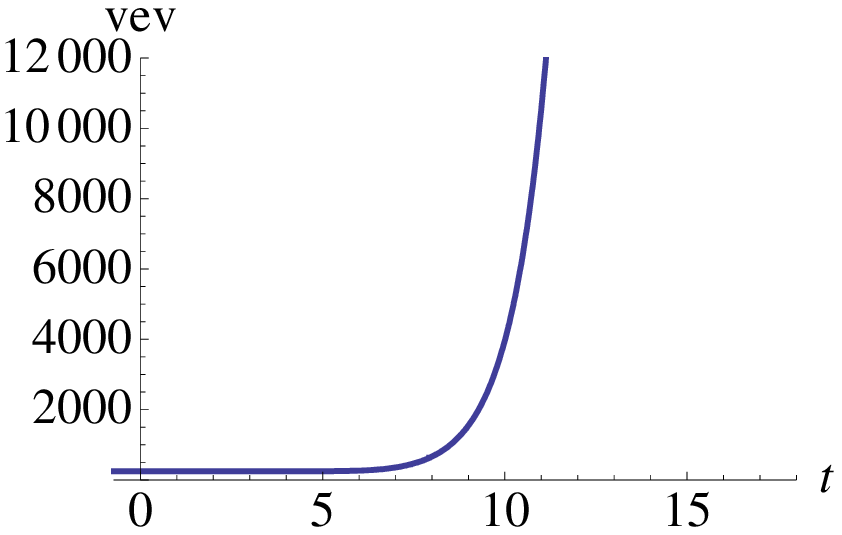}
 \includegraphics[width=0.3\textwidth]{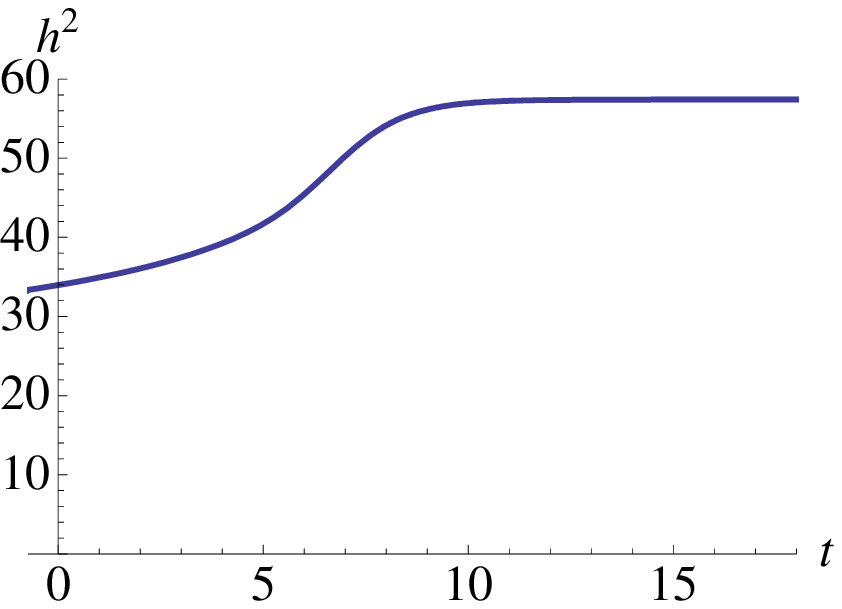}
 \includegraphics[width=0.3\textwidth]{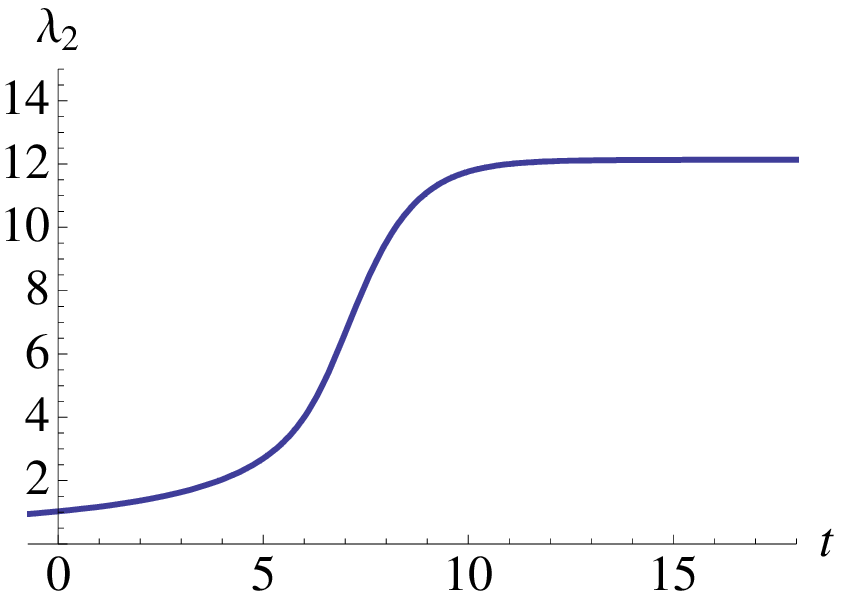}
 \caption{Flow of the leading couplings within the lowest-order truncation.}
\label{fig:ssbflow}
\end{center}
\end{figure}

This counting of independent physical parameters changes drastically if the
system originates from the non-Gau\ss ian fixed point in the limit of an
infinite UV cutoff. As the non-Gau\ss ian fixed point has only one relevant
direction, there is only one physical parameter which we relate to the vev
$v=246$GeV. In fact, fixing $v$ to its physical value merely corresponds to
fixing the absolute scale. Dimensionless couplings or mass ratios are even
independent of this fixing scale and thus are completely parameter-free
predictions of the theory. 

In practice, we have to start the flow at some finite value of the auxiliary
UV cutoff $\Lambda$ (in arbitrary units), solve the flow, and then associate
the resulting value of $v$ with its correct value in physical units. This also
fixes the value of $\Lambda$ in units of, say, GeV.  In the model with
$N_\mathrm{L}=10$, we then end up with a top and a Higgs mass at the IR scale
\begin{equation}\label{eq:topmass2}
m_{\text{top}}= 5.78 v, \quad  m_{\text{Higgs}}= 0.97 v,
\end{equation}
where the proportionality factors are approached in the limit $v/\Lambda\ll1$.
In this limit, this result is robust as we vary the UV cutoff, which
demonstrates the cutoff independence of our prediction. As a result, the top
and the Higgs mass are pure predictions of the theory itself. 

Whereas the Higgs mass is of the order of the vev as expected,
$m_{\text{Higgs}}\simeq 239$GeV for $v=246$GeV, the top mass is significantly
larger in our model: $m_{\text{top}}\simeq 1422$GeV. This is an indirect
consequence of the comparatively large value of the fixed point of the Yukawa
coupling. In order to elucidate this relation, let us note that the flow can
be subdivided into three different regimes: first, the UV fixed-point regime
where all couplings stay close to their fixed point values; second, a
crossover region where all couplings run fast; and third, a freeze-out region
in the IR where the dimensionful vev builds up and induces decoupling and a
subsequent approach to the generically frozen IR values. 

The dynamics in the crossover region is an intrinsic parameter-free property
of the model that occurs only in a rather narrow window of momentum
scales. Hence, the couplings have only a finite ``RG time'' -- similar to the
number of $e$-foldings in inflationary cosmology -- to run from their UV
initial conditions to their IR-frozen values. As the initial fixed-point
Yukawa coupling in our model is large, a sizeable IR value remains and gives
rise to a large top mass. The same mechanism has been identified in the
Z${}_2$-symmetric Yukawa system in \cite{GiesScherer:2009} even for the case
with two relevant directions. There, the top mass is, in principle, a physical
parameter to be fixed; still, the finite number of crossover $e$-foldings
leads to a lower bound on the Higgs mass.

Realistic scenarios with a physical value of the top mass require either a
smaller fixed-point Yukawa coupling, or a sufficiently large
number of crossover $e$-foldings that facilitates a long running of the Yukawa
coupling to its physical value. As the number of $e$-foldings of the crossover
window is mainly dictated by the RG speed at which the physical vev builds up,
large numbers of $e$-foldings can occur if the largest critical exponent
dominating the $\kappa$ flow is small. We conclude that a reduced hierarchy
problem and a realistic value of the top mass can go hand in hand with each other.

Finally, we have to address an artifact of our flow in the IR. As can be read
off from Fig.~\ref{fig:ssbflow}, the perturbatively marginal couplings $h^2$
and $\lambda_2$ still exhibit a slow log-like running in the IR. In our
present set of flow equations, the source of this running can be traced back
to the Goldstone modes which are a particularity of our model, but are not
present in the standard model. However, even if we considered the Goldstone
modes as physical, the log-like running is still an artifact of our truncation
in the IR. This artifact arises from the Cartesian decomposition of the scalar
field, $\phi = \frac{1}{\sqrt{2}}( \phi_1+i\phi_2)$ (we suppress flavor
indices in the following argument) where the vev is associated with the
$\phi_1$ direction and $\phi_2$ is populated with Goldstone modes. This
decomposition leads to log-like divergencies in the diagrams with internal
$\phi_2$ loops coupling to external $\phi_1$ legs. However, as becomes
clear from a proper nonlinear decomposition of the scalar field
$\phi=\sqrt{\rho} e^{i\theta}$, the true Goldstone modes encoded in the field
$\theta$ do not couple to the radial mode via the potential, as
$U(\phi^\dagger \phi)=U(\rho)$ is independent of $\theta$. We conclude
that the log-like running would be absent in a proper nonlinear field basis in
the broken regime \cite{PawlowskiLamp:2009}.

\begin{figure}
\begin{center}
 \includegraphics[width=0.3\textwidth]{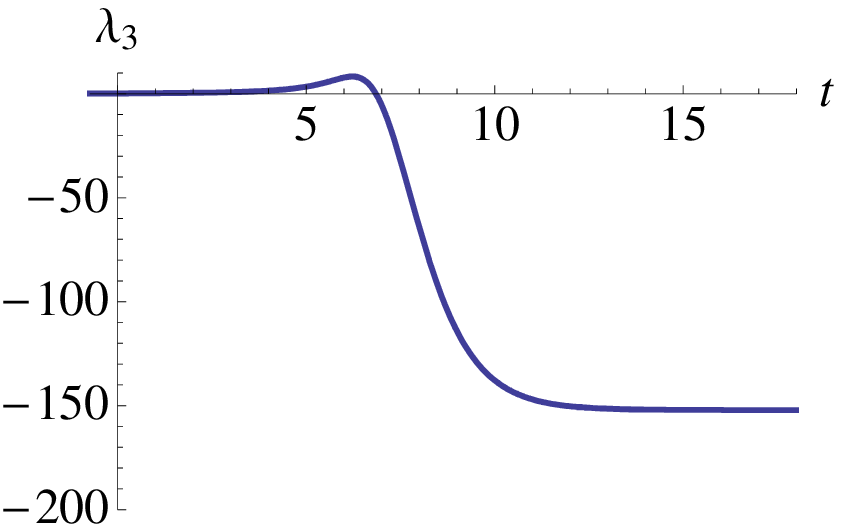}
 \includegraphics[width=0.3\textwidth]{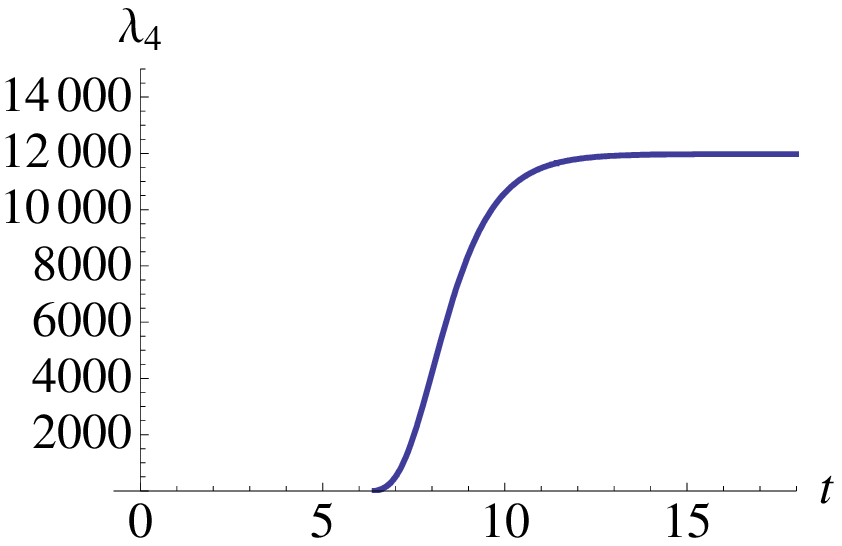}
 \caption{Flow of the higher-order couplings $\lambda_3$ and $\lambda_4$ of the effective potential. The flows of even higher order couplings $\lambda_5$ and $\lambda_5$ look similar.}
\label{fig:ssbflow2}
\end{center}
\end{figure}

In practice, we read off our estimates for the IR values of $h^2$ and
$\lambda_2$ at a dynamically defined scale $k_{\text{IR}}$. For the
identification of this scale, we note that higher-order couplings
$\lambda_{n>2}$ are not affected by the Cartesian basis and safely flow to
zero due to the leading-order Gau\ss ian flow, 
\begin{equation}
 \partial_t \lambda_n = 2(n-2)\lambda_n+...\,.
\end{equation}
as expected, see Fig.~\ref{fig:ssbflow2}. We define the scale
$k_{\text{IR}}$ by that scale at which the coupling $\lambda_{3}$ approaches
its zero IR fixed point from above within an accuracy of $0.1\%$. The IR
predictions listed in \Eqref{eq:topmass2} have been read off at this scale. Of
course, in the presence of this flow artifact, these predictions actually
depend on the choice of $k_{\text{IR}}$. Nevertheless, the generic mechanisms
in the UV and crossover regimes outlined above are not affected by the IR
artifact.

\section{Properties of the derivative expansion at next-to-leading order}
\label{sec:NLO}

In the preceeding section, we have analyzed the chiral Yukawa model in a
derivative expansion of the effective action. As a simple criterion for this
expansion to be valid, the anomalous dimensions measuring the influence of
higher-derivative terms should be small, cf. \Eqref{eq:validity}. 

\begin{figure}
\begin{center}
 \includegraphics[width=0.3\textwidth]{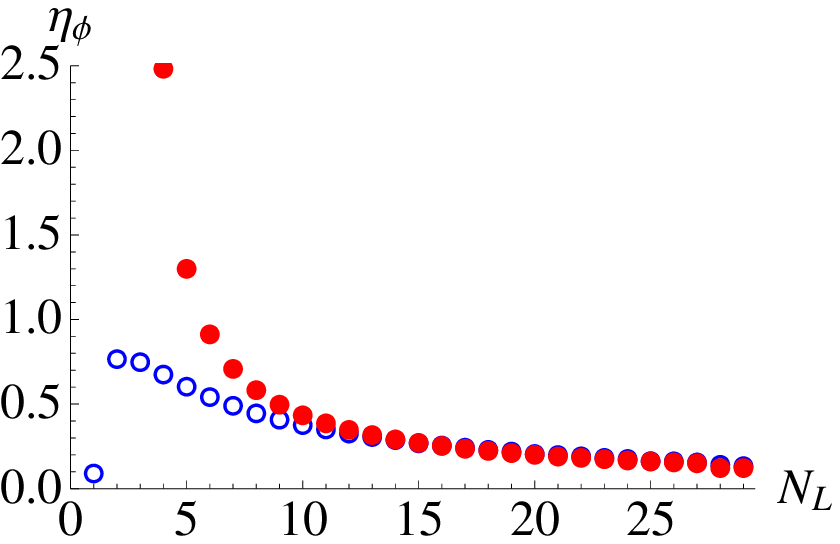}
 \includegraphics[width=0.3\textwidth]{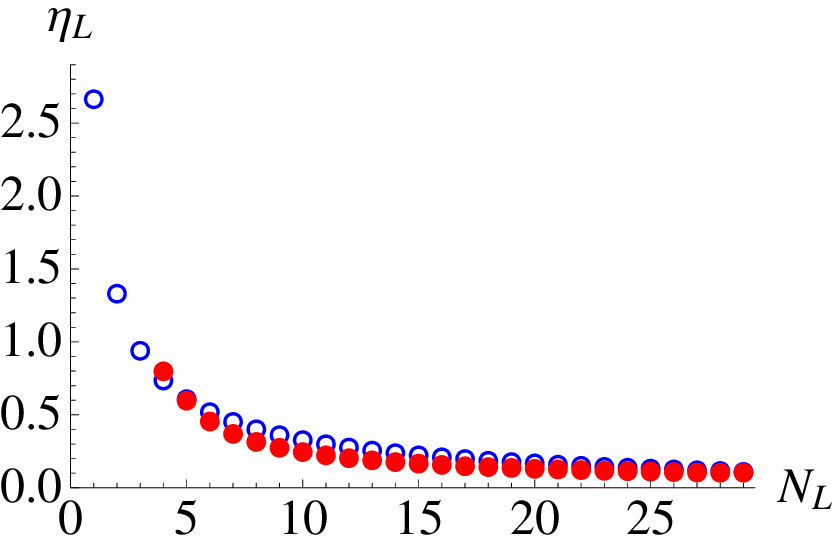}
 \includegraphics[width=0.3\textwidth]{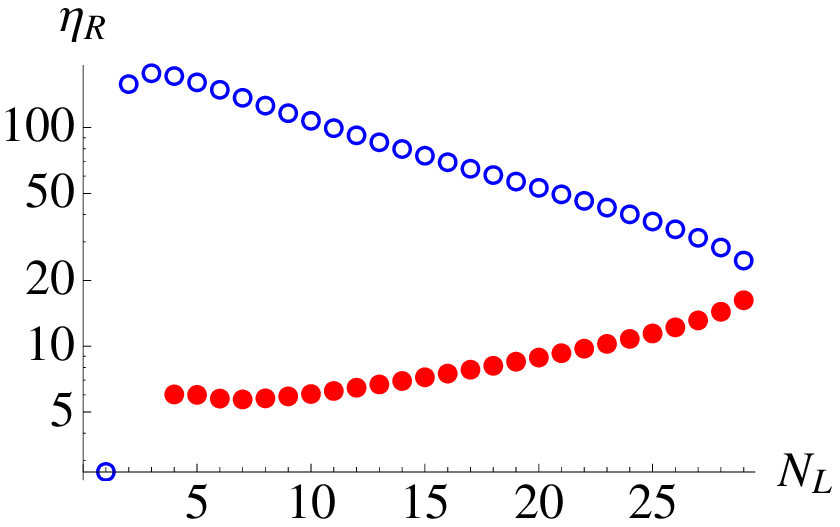}
 \caption{Leading-order estimate of the anomalous dimensions.}
\label{fig:ssbanom}
\end{center}
\end{figure}

As a first check of this criterion, we use our fixed-point results from the
lowest-order polynomial expansion in the local-potential approximation and
insert them into the right-hand sides of the anomalous dimensions
Eqs.~\eqref{eq:etaphi}-\eqref{eq:etaR}. For this estimate, we ignore the RG
improvement in the form of the back-reactions of nonzero $\eta$'s on the flow
of the couplings. The results as a function of $\NL$ are shown in
Fig.~\ref{fig:ssbanom}. Whereas the criterion \eqref{eq:validity} is satisfied
for $\eta_\phi$ and $\eta_\text{L}$ for $\NL>5$, the anomalous dimension
$\eta_{\text{R}}$ of the right-handed fermion violates the criterion for all
$\NL$. The violation is dramatic for the fixed point with two relevant
directions and still substantial for the fixed point with one relevant
direction.

One reason for the size of $\eta_{\text{R}}$ lies in the fact that the
massless Goldstone modes and massless bottom-type fermions contribute
strongly. This is because they (i) are not damped by massive threshold effects
induced by couplings to the condensate, and (ii) contribute with a large
multiplicity $\sim\NL$. We conclude that the leading-order derivative
expansion does not provide for self-consistent estimates of the fixed-point
structure of the chiral Yukawa model as such. Nevertheless, since the
Goldstone as well as the massless fermion modes are not present in the
standard model but a particularity of our reduced toy model, we expect that
the local-potential approximation of the present model provides for a better
picture of a possible fixed-point structure of the standard-model Higgs sector
than the literal version of our model including its massless modes.

It is still an interesting question within the chiral Yukawa model, whether
the nonlinearities induced by the anomalous dimension gives rise to admissible
fixed points if the full RG improvement is taken into account. For this, we
have first followed the evolution of our leading-order fixed points upon
gradually switching on the anomalous dimensions. Our numerical results are
compatible with a destabilization of this fixed point, potentially going along
with an annihilation with other fixed points. A numerical search of further
admissible fixed points upon the full inclusion of the anomalous dimension
remained inconclusive. Owing to the high degree of nonlinearity of the coupled
equations, a systematic search in this high-dimensional parameter space is
computationally expensive and beyond the scope of the present work.

\section{Conclusions}
\label{sec:conc}

We have explored a new possible route to asymptotically safe Yukawa systems
with a chiral U$(N_\mathrm{L})_\mathrm{L}\otimes$U$(1)_\mathrm{R}$ symmetry,
similar to the Higgs-top sector of the standard model. This work is motivated
by the fact that an asymptotically safe Yukawa sector can solve the triviality
problem of the standard-model Higgs sector and potentially improve on the
hierarchy problem, both of which represent genuine problems related to the UV
stability and completion of standard-model particle physics. In our scenario
proposed in \cite{GiesScherer:2009}, asymptotic safety is based on a conformal
Higgs expectation value that induces a non-Gau\ss ian UV fixed point for all
essential couplings. 

This conformal behavior can arise from the dynamics of the theory as a
consequence of a balancing between fermionic and bosonic fluctuation
contributions to the running of the vev. More specifically, since bosonic
fluctuations need to dominate, the existence of this conformal-vev behavior
depends on the algebraic structure of the theory. It is a central result of
our work that chiral models with an asymmetry between left- and right-handed
fermions in fact support this boson dominance.

Our quantitative investigations are based on the functional RG, using the
derivative expansion as a nonperturbative approximation scheme for a
systematic calculation of the quantum effective action. To leading-order
(local-potential approximation), our left-right asymmetric model exhibits the
desired non-Gau\ss ian fixed points for $1\leq \NL \leq 57$. Whereas a
physically admissible non-Gau\ss ian fixed point establishes the existence of
the Yukawa model as a fundamental UV complete quantum field theory, the
physical properties of such a UV completion are quantified by the non-negative
critical exponents of this fixed point. Their number corresponds to the number
of physical parameters, and their value is a measure of the naturalness of a
large hierarchy (between, say, the Planck and the electroweak scale). 

Most importantly, one of the admissible fixed points has only one positive
critical exponent. This implies that only one physical parameter has to be fixed,
e.g., the vev $v=246$GeV, whereas all other IR quantities such as the Higgs or
the top mass are a pure prediction of the theory. Moreover, this critical
exponent is also smaller than the corresponding value in the perturbative
domain, such that the hierarchy behavior is substantially improved. 

Unfortunately, the fixed point in our toy model is destabilized at higher
order in the derivative expansion due to massless Goldstone and fermion
fluctuations. As the latter are a particularity of our model and are not
present in the standard model, we expect the results from the local-potential
approximation to be of high relevance for a more realistic model of the
standard-model Higgs sector. On the other hand, our results have proved robust
under an extension of the approximation scheme in the boson effective
potential.

As a main result of the present study, a number of conclusions about the
requirements for a more realistic model can be drawn. (1.) The appearance of
the conformal-vev mechanism depends crucially on the algebraic structure of
the theory. Boson dominance is required and can be induced by a chiral
left-right asymmetry. (2.) If the conformal vev breaks a continuous global
symmetry, Goldstone bosons are generated. As they are not controlled by
threshold effects, they may have the tendency to destabilize the UV fixed
point. A model with spontaneous breaking of gauged symmetries will not be
affected by this problem. (3.) Even though the threshold effects of the
conformal vev are nonperturbative, our mechanism does technically not require
strong coupling. Comparatively small UV fixed-point couplings would also be
allowed if not phenomenologically preferred. (4.) The present model predicts a
very large top mass. For a realistic top mass, either the Yukawa UV
fixed-point value should not be too large or the crossover from the UV regime
to the IR freeze-out should extend over many ``$e$-foldings'', i.e., a larger
range of RG time.

These requirements together with the remarkable predictive power of our
scenario are a strong motivation to extend the present set of ideas to gauged
chiral Yukawa models, thereby making another step towards a realistic standard-model
Higgs sector.

\acknowledgments

The authors are grateful to J. M. Pawlowski for interesting discussions and helpful
comments. This work was supported by the DFG under contract No. Gi 328/5-1
(Heisenberg program) and FOR 723.

\appendix

\section{Threshold functions}\label{section:threshold}

The boson cutoff we use is given by:
\begin{equation}\label{eq:cutoff}
y r_{\mathrm{B}}(y)=(1-y)\theta(1-y),
\end{equation}
where $y=q^2/k^2$, and the fermion regulator $r_{\mathrm{F}}(y)$ is chosen such that $y(1+r_{\mathrm{B}})=y(1+r_{\mathrm{F}})^2$. Using this regulator in the Wetterich equation, we can perform all momentum integrations analytically. The result can be given in terms of threshold functions which read
\begin{eqnarray}
l^d_n(\omega) &=& \frac{2(\delta_{n,0}+n)}{d}\Big(1
-\frac{\eta_\phi}{d+2}\Big) \frac{1}{(1+\omega)^{n+1}}, \label{eq:threshold1}\nonumber\\
l^{(\mathrm{F})d}_{n,\mathrm{L/R}}(\omega) &=& \frac{2(\delta_{n,0}+n)}{d}\Big(1
-\frac{\eta_{\mathrm{L/R}}}{d+1}\Big) \frac{1}{(1+\omega)^{n+1}},\nonumber\\
l^{(\mathrm{FB})d}_{n_1,n_2}(\omega_1,\omega_2) &=&
\frac{2}{d}\frac{1}{(1+\omega_1)^{n_1}(1+\omega_2)^{n_2}}\nonumber\\
&\times&\left\{\frac{n_1\left(1-\frac{\eta_\psi}{d+1}\right)}{1+\omega_1}
 + \frac{n_2\left(1-\frac{\eta_\phi}{d+2}\right)}{1+\omega_2}\right\},\nonumber\\
m^d_{n_1,n_2}(\omega_1,\omega_2)
  &=& \frac{1}{(1+\omega)^{n_1}(1+\omega)^{n_2}},\label{eq:thresholdFunctions}\\
m^{(\mathrm{F})d}_{2}(\omega) &=& \frac{1}{(1+\omega)^4}\quad,\nonumber\\
m^{(\mathrm{F})d}_{4}(\omega) &=& \frac{1}{(1+\omega)^4}+\frac{1-\eta_\psi}{d-2}\frac{1}{(1+\omega)^3}\nonumber\\
&{}& - \left(\frac{1-\eta_\psi}{2d-4}+\frac{1}{4}\right)\frac{1}{(1+\omega)^2},\nonumber\\
m^{(\mathrm{FB})d}_{n_1,n_2}(\omega_1,\omega_2)
 &=& \left(1-\frac{\eta_\phi}{d+1}\right)\frac{1}{(1+\omega_1)^{n_1}(1+\omega_2)^{n_2}}\,, \label{eq:threshold2}\nonumber
\end{eqnarray}
where we have defined $\eta_\psi:=\frac{1}{2}(\eta_{\mathrm{R}}+\eta_{\mathrm{L}})$.

\section{Derivation of the effective potential flow}
\label{sec:deriv-effect-potent}

For the flow of the effective potential, we project the Wetterich equation onto constant bosonic fields and vanishing fermionic fields,
\begin{equation}\label{eq:PotentialFlow}
\partial_tU_k=\frac{1}{2\Omega}\mathrm{STr}\{(\Gamma^{(2)}_k+R_k)^{-1}(\partial_tR_k)\}|_{\phi=\mathrm{const.}, \psi=0},
\end{equation}
where $\Omega$ is the spacetime volume. We have to evaluate the r.h.s of \Eqref{eq:PotentialFlow}, for which we need the $\Gamma_k^{(2)}$ matrix. Taking care of the partly Grassmann-valued field components and the Fourier conventions, $\Gamma_k^{(2)}$ is derived by
\begin{equation*}
 \Gamma_k^{(2)}=
\begin{pmatrix}
\frac{\overrightarrow{\delta}}{\delta \phi_1(-p)}\\
\frac{\overrightarrow{\delta}}{\delta \phi_2(-p)}\\
\frac{\overrightarrow{\delta}}{\delta \psi_{\mathrm{L}}(-p)}\\
\frac{\overrightarrow{\delta}}{\delta \bar{\psi}_{\mathrm{L}}^{\mathrm{T}}(p)}\\
\frac{\overrightarrow{\delta}}{\delta \psi_{\mathrm{R}}(-p)}\\
\frac{\overrightarrow{\delta}}{\delta \bar{\psi}_{\mathrm{R}}^{\mathrm{T}}(p)}
\end{pmatrix}^{\mathrm{T}}
\Gamma_k
\begin{pmatrix}
\frac{\overleftarrow{\delta}}{\delta \phi_1(q)} \\
\frac{\overleftarrow{\delta}}{\delta \phi_2(q)} \\
\frac{\overleftarrow{\delta}}{\delta \psi_{\mathrm{L}}(q)} \\
\frac{\overleftarrow{\delta}}{\delta \bar{\psi}_{\mathrm{L}}^{\mathrm{T}}(-q)} \\
\frac{\overleftarrow{\delta}}{\delta \psi_{\mathrm{R}}(q)} \\
\frac{\overleftarrow{\delta}}{\delta \bar{\psi}_{\mathrm{R}}^{\mathrm{T}}(-q)}
\end{pmatrix},
\end{equation*}
with
\begin{eqnarray*}
\phi_i^{\mathrm{T}}&=&\begin{pmatrix} \phi_i^{1\mathrm{T}},\ldots, \phi_i^{N_{\mathrm{L}}\mathrm{T}} \end{pmatrix},\\
\bar{\psi}_{\mathrm{L}}&=&\begin{pmatrix} \bar{\psi}_{\mathrm{L}}^{1},\ldots, \bar{\psi}_{\mathrm{L}}^{N_{\mathrm{L}}} \end{pmatrix}.
\end{eqnarray*}
The transposition refers to flavor as well as Dirac indices. For a proper IR regularization, a regulator which is diagonal in field space is sufficient and convenient,
\begin{equation*}
R_k(q,p)=\delta(p-q)
\begin{pmatrix}
R_{k\mathrm{B}} & 0\\
0 & R_{k\mathrm{F}}
\end{pmatrix},
\end{equation*}
with a $2N_\mathrm{L}\times 2N_\mathrm{L}$ matrix for the bosonic sector
\begin{equation*}
R_{k\mathrm{B}}=
\begin{pmatrix}
Z_{\phi , k}\delta^{ab}p^2r_{\mathrm{B}} & 0\\
0 & Z_{\phi , k}\delta^{ab}p^2r_{\mathrm{B}}
\end{pmatrix},
\end{equation*}
cf. App. \ref{section:threshold} and an $(2N_\mathrm{L}+2)\times (2N_\mathrm{L}+2)$ matrix for the fermionic sector
%
%\begin{widetext}
\begin{equation*}
R_{k\mathrm{F}}=
-\begin{pmatrix}
0 & Z_{\mathrm{L},k}\delta^{ab}\pslash^{\mathrm{T}} & 0 & 0\\
Z_{\mathrm{L},k}\delta^{ab}\pslash & 0 & 0 & 0\\
0 & 0 & 0 & Z_{\mathrm{R},k}\pslash^{\mathrm{T}}\\
0 & 0 & Z_{\mathrm{R},k}\pslash & 0
\end{pmatrix}r_{\mathrm{F}}.
\end{equation*}
%\end{widetext}
%
The matrix $\Gamma_k^{(2)}+R_k$ has the same block form as the regulator. Therefore, we can evaluate the bosonic and fermionic parts separately. We start with the bosonic part. Inverting the matrix, multiplying with the derivative of the regulator, and taking the supertrace yields
\begin{eqnarray*}
\partial_tU_{k\mathrm{B}}&=&\frac{1}{2}\int \frac{d^dp}{(2\pi)^d}\partial_tR_k\Big[ \frac{2N_{\mathrm{L}}-1}{Z_{\phi ,k}P_{\mathrm{B}}(p)+U_k'}\\
&{}&+\frac{1}{Z_{\phi ,k}P_{\mathrm{B}}(p)+U_k'+2U_k''\rho} \Big],
\end{eqnarray*}
where $P_{\mathrm{B}}(p)=p^2(1+r_{\mathrm{B}}(p))$. Introducing $P_{\mathrm{F}}(p)=p^2(1+r_{\mathrm{F}}(p))^2$ and $d_{\gamma}$ as the dimension of the representation of the Dirac algebra, the fermionic contribution reads
\begin{eqnarray*}
\partial_tU_{k\mathrm{F}}&=&-d_{\gamma}\int\frac{d^dp}{(2\pi)^d}\Big\{\frac{\partial_t[Z_{\mathrm{L},k}r_{\mathrm{F}}(p)]}{Z_{\mathrm{L},k}(1+r_{\mathrm{F}}(p))} \\
&{}&\times\Big[(N_{\mathrm{L}}-1)+\frac{Z_{\mathrm{L},k}Z_{\mathrm{R},k}P_{\mathrm{F}}(p)}{\bar{h}_k^2\rho+Z_{\mathrm{L},k}Z_{\mathrm{R},k}P_{\mathrm{F}}(p)}\Big] \\ &{}&+\frac{\partial_t[Z_{\mathrm{R},k}r_{\mathrm{F}}(p)]}{Z_{\mathrm{R},k}(1+r_{\mathrm{F}}(p))}\frac{Z_{\mathrm{L},k}Z_{\mathrm{R},k}P_{\mathrm{F}}(p)}{\bar{h}_k^2\rho+Z_{\mathrm{L},k}Z_{\mathrm{R},k}P_{\mathrm{F}}(p)}\Big\}.
\end{eqnarray*}
Adding both parts and using the optimized regulator introduced in App. \ref{section:threshold}, we get
\begin{eqnarray*}\label{eq:PotentialFlowThreshold}
\partial_tU_k&=&2v_dk^d\Big[(2N_{\mathrm{L}}-1)l_0^d\Big(\frac{U_k'}{Z_{\phi,k}k^2}\Big)+l_0^d\Big(\frac{U_k'+2U_k''\rho}{Z_{\phi,k}k^2}\Big)\Big]\\
&{}&-d_{\gamma}v_dk^d2\Big[ (N_{\mathrm{L}}-1)l_{0,\mathrm{L}}^{(\mathrm{F})d}(0)\\
&{}&+l_{0,\mathrm{L}}^{(\mathrm{F})d}\Big(\frac{\bar{h}_k^2\rho}{k^2Z_{\mathrm{L},k}Z_{\mathrm{R},k}}\Big)+l_{0,\mathrm{R}}^{(\mathrm{F})d}\Big(\frac{\bar{h}_k^2\rho}{k^2Z_{\mathrm{L},k}Z_{\mathrm{R},k}}\Big) \Big],
\end{eqnarray*}
where the threshold functions are defined in
Eqs.\eqref{eq:thresholdFunctions}. In terms of dimensionless quantities
defined in \Eqref{eq:dimensionless}, the potential flow turns into
Eq. \eqref{basic:flowequation} in the main text.\bigskip

\section{Derivation of the Yukawa coupling flow}\label{section:yukawaflow}

For the derivation of the flow of the Yukawa coupling, we first separate the bosonic field into a vacuum expectation value (vev) $v$ and a purely radial deviation from the vev, since we are mainly interested in the Yukawa coupling between the fermions and the radial mode.
\begin{eqnarray*}
\phi(p)&=&\frac{1}{\sqrt{2}}
\begin{pmatrix}
\phi_1^{1}(p)+i\phi_2^{1}(p)\\
\phi_1^{2}(p)+i\phi_2^{2}(p)\\
\vdots\\
\phi_1^{N_\mathrm{L}}(p)+i\phi_2^{N_\mathrm{L}}(p)
\end{pmatrix}
\\
&=&\frac{1}{\sqrt{2}}
\begin{pmatrix}
v\\
0\\
\vdots\\
0
\end{pmatrix}\delta(p)+\frac{1}{\sqrt{2}}
\begin{pmatrix}
\Delta\phi_1^{1}(p)\\
\Delta\phi_1^{2}(p)\\
\vdots\\
\Delta\phi_1^{N_\mathrm{L}}(p)
\end{pmatrix},
\end{eqnarray*}
setting all $\Delta\phi_2$ Goldstone components to zero. The projection of the Wetterich equation onto the flow of the Yukawa coupling reads
\begin{equation}\label{eq:flowh}
\partial_t\bar{h}_k=\frac{-1}{2}\frac{\overrightarrow{\delta}}{\delta\bar{\psi}_{\mathrm{L}}^1(p)}\frac{\sqrt{2}\overrightarrow{\delta}}{\delta\Delta\phi_1^{1}(p')} \partial_t\Gamma_k \frac{\overleftarrow{\delta}}{\delta\psi_{\mathrm{R}}(q)}\Bigg|.
\end{equation}
The vertical line indicates that the equation is evaluated at $\psi_{\mathrm{R}}^a=\psi_{\mathrm{L}}^a=\Delta\phi=0,\ p'=p=q=0$. Next, we can decompose the matrix $(\Gamma_k^{(2)}+R_k)$ into two parts. One part, which we call $(\Gamma_{k,0}^{(2)}+R_k)$, contains only $v$ and is independent of the fluctuations. The remaining part, $\Delta \Gamma_k^{(2)}$, contains all fluctuating fields. Inserting this into equation \eqref{eq:flowh} and expanding the logarithm
\begin{widetext}
\begin{eqnarray}
\mathrm{STr} \left( \mathrm{ln}(\Gamma_k^{(2)}+R_k)\right)&=&\mathrm{STr}\left( \mathrm{ln}\left[ (\Gamma_{k,0}^{(2)}+R_k)\left(1+\frac{\Delta \Gamma_k^{(2)}}{\Gamma_{k,0}^{(2)}+R_k}\right) \right]\right)\\
&=&\mathrm{STr}\left( \mathrm{ln}(\Gamma_{k,0}^{(2)}+R_k)\right)+\mathrm{STr}\frac{\Delta \Gamma_k^{(2)}}{\Gamma_{k,0}^{(2)}+R_k}-\frac{1}{2}\mathrm{STr}\left( \frac{\Delta \Gamma_k^{(2)}}{\Gamma_{k,0}^{(2)}+R_k} \right)^2+\frac{1}{3}\mathrm{STr}\left( \frac{\Delta \Gamma_k^{(2)}}{\Gamma_{k,0}^{(2)}+R_k} \right)^3-\ldots,\label{eq:logarithm}\nonumber
\end{eqnarray}
only the term to third power in $\Delta\Gamma_k^{(2)}$ survives the projection onto $\Delta\phi_1^1 \bar{\psi}_{\mathrm{L}}^1 \psi_{\mathrm{R}}$. Performing the matrix calculations and taking the supertrace, we get
\begin{eqnarray*}
\partial_t\bar{h}_k=-\frac{\bar{h}_k^3}{2}\int\frac{d^dp}{(2\pi)^d}\tilde{\partial}_t
\left[\frac{v}{Z_{\mathrm{L},k}Z_{\mathrm{R},k}P_{\mathrm{F}}(p)+\frac{\bar{h}_k^2}{2}v^2}
\left(\frac{U_k''v}{(Z_{\phi ,k}P_{\mathrm{B}}(p)+U_{k}')^2}
-\frac{3U_{k}''v+U_{k}'''v^3}{(Z_{\phi ,k}P_{\mathrm{B}}(p)+U_{k}'+U_{k}''v^2)^2}\right)\right.\\
+\frac{\bar{h}_k^2v^2}{(Z_{\mathrm{L},k}Z_{\mathrm{R},k}P_{\mathrm{F}}(p)+\frac{\bar{h}_k^2}{2}v^2)^2}
\left(\frac{1}{Z_{\phi ,k}P_{\mathrm{B}}(p)+U_{k}'}-\frac{1}{Z_{\phi ,k}P_{\mathrm{B}}(p)+U_{k}'+U_{k}''v^2}\right)\\
\left.-\frac{1}{Z_{\mathrm{L},k}Z_{\mathrm{R},k}P_{\mathrm{F}}(p)+\frac{\bar{h}_k^2}{2}v^2}
\left(\frac{1}{Z_{\phi ,k}P_{\mathrm{B}}(p)+U_{k}'}-\frac{1}{Z_{\phi ,k}P_{\mathrm{B}}(p)+U_{k}'+U_{k}''v^2}\right)\right],
\end{eqnarray*}
where the potential on the right hand side is evaluated at the minimum $\frac{1}{2}v^2$. Using the optimized regulator and the threshold functions as defined in App. \ref{section:threshold} and switching over to dimensionless quantities, we end up with the representation \eqref{eq:flowh2} given in the main text.
\end{widetext}

\section{Derivation of the anomalous dimensions}

For the derivation of the flow of $Z_{\phi ,k}$, we decompose the bosonic field as in App. \ref{section:yukawaflow}. The projection of the Wetterich equation onto the boson kinetic term leads us to
\begin{eqnarray} \label{eq:etaphiflow}
 \partial_t Z_{\phi ,k}&=&-\frac{\partial}{\partial(p'^2)}\frac{\delta}{\delta \Delta\phi_1^1(p')}\frac{\delta}{\delta \Delta\phi_1^1(q')}\\
&{}&\times\frac{1}{4}\mathrm{STr}\left[ \tilde{\partial}_t\left( \frac{\Delta\Gamma_k^{(2)}}{\Gamma_k^{(2)}+R_k} \right)^2 \right]\Bigg|_{\Delta\phi=\psi_{\mathrm{L}}^a=\psi_{\mathrm{R}}=0, p'=q'=0}.\nonumber
\end{eqnarray}
Again, we have decomposed the matrix into two parts, one part containing the vev and the other part containing all fluctuating fields.  As done in App. \ref{section:yukawaflow}, we have expanded the logarithm as in \Eqref{eq:logarithm}, but this time only the second order contributes. Calculating the r.h.s of \Eqref{eq:etaphiflow}, results in
\begin{widetext}
\begin{eqnarray*}
\partial_tZ_{\phi , k}&=&\frac{1}{d}\int\frac{d^dp}{(2\pi)^d}\tilde{\partial}_t\Big[(3U_{k}''v+U_{k}'''v^3)^2p^2Z_{\phi , k}^2\left( \frac{\frac{\partial}{\partial p^2}P_{\mathrm{B}}(p)}{(Z_{\phi , k}P_{\mathrm{B}}(p)+U_{k}'+U_{k}''v^2)^2} \right)^2\\
&&+(2N_{\mathrm{L}}-1)(U_{k}''v)^2p^2Z_{\phi , k}^2\left( \frac{\frac{\partial}{\partial p^2}P_{\mathrm{B}}(p)}{(Z_{\phi , k}P_{\mathrm{B}}(p)+U_{k}')^2} \right)^2\\
&&+2h_k^2d_{\gamma}p^4Z_{\mathrm{L},k}Z_{\mathrm{R},k}\left( \frac{\partial}{\partial p^2}\frac{(1+r_{\mathrm{F}}(p))}{Z_{\mathrm{L}, k}Z_{\mathrm{R},k}P_{\mathrm{F}}(p)+\frac{h_k^2}{2}v^2} \right)^2-h_k^4d_{\gamma}v^2p^2\left( \frac{\partial}{\partial p^2}\frac{1}{Z_{\mathrm{L},k}Z_{\mathrm{R},k}P_{\mathrm{F}}(p)+\frac{h_k^2}{2}v^2} \right)^2 \Big],
\end{eqnarray*}
where the potential is evaluated at the minimum. Using the optimized regulator together with the threshold functions as defined in App. \ref{section:threshold} and again introducing dimensionless quantities, we end up with
\begin{eqnarray}
\eta_{\phi}&=&\frac{8 v_d}{d}\tilde\rho(3 u_k'' +2\tilde\rho u_k''')^2m_{22}^d(u_k'+2\tilde\rho u_k'')+\frac{(2N_{\mathrm{L}}-1)8v_d}{d}\tilde\rho u_k''^2 m_{22}^d(u_k')\nonumber\\
&&+\frac{8v_d d_\gamma}{d}h_k^2 m_4^{(\mathrm{F})4}(\tilde\rho h_k^2)-\frac{8v_d d_\gamma}{d}\tilde\rho h_k^4m_2^{(\mathrm{F})4}(\tilde\rho h_k^2),\nonumber
\end{eqnarray}
where we have used $\eta_{\phi}=-\frac{\partial_t Z_{\phi,k}}{Z_{\phi,k}}$, and the right hand side has to be evaluated on the vev.

For the fermionic anomalous dimensions, the procedure is the same. We start with
\begin{equation}
\partial_tZ_{\mathrm{L/R},k}=\left.\frac{1}{4v_dd_{\gamma}}\mathrm{tr}\gamma^{\mu}\frac{\partial}{\partial p'^{\mu}}\frac{\overrightarrow{\delta}}{\delta\bar{\psi}_{\mathrm{L/R}}^1(p')}\mathrm{STr}\left[ \left( \frac{\Delta\Gamma_k^{(2)}}{\Gamma_{k,0}^{(2)}+R_k} \right)^2 \right]\frac{\overleftarrow{\delta}}{\delta\psi_{\mathrm{L/R}}^1(q')}\right|_{\Delta\phi=\psi_{\mathrm{L}}=\psi_{\mathrm{R}}=0,p'=q'=0},
\end{equation}
and get
\begin{eqnarray*}
\partial_tZ_{\mathrm{L},k}&=&\frac{2\bar{h}_k^2}{d}\int \frac{d^dp}{(2\pi)^d}p^2\tilde{\partial}_t\Big[ \frac{Z_{\mathrm{L},k}(1+r_{\mathrm{F}}(p))}{Z_{\mathrm{L},k}Z_{\mathrm{R},k}P_{\mathrm{F}}(p)+\frac{\bar{h}_k^2}{2}v^2}Z_{\phi , k}\frac{\partial}{\partial p^2}P_{\mathrm{B}}(p)\Big( \frac{1}{(Z_{\phi , k}P_{\mathrm{B}}(p)+U_{k}'+U_{k}''v^2)^2}\\
&{}&+\frac{1}{(Z_{\phi , k}P_{\mathrm{B}}(p)+U_{k}')^2} \Big) \Big]
\end{eqnarray*}
and
\begin{eqnarray*}
\partial_tZ_{\mathrm{R},k}&=&\frac{2\bar{h}_k^2}{d}\int \frac{d^dp}{(2\pi)^d}p^2\tilde{\partial}_t
\Big[ \frac{Z_{\mathrm{R},k}(1+r_{\mathrm{F}}(p))}{Z_{\mathrm{L},k}Z_{\mathrm{R},k}P_{\mathrm{F}}(p)+\frac{\bar{h}_k^2}{2}v^2}Z_{\phi , k}\frac{\partial}{\partial p^2}P_{\mathrm{B}}(p)\left( \frac{1}{(Z_{\phi , k}P_{\mathrm{B}}(p)+U_{k}'+U_{k}''v^2)^2} \right)\\
&{}&+\frac{1}{(Z_{\phi , k}P_{\mathrm{B}}(p)+U_{k}')^2}+ \frac{2(N_{\mathrm{L}}-1)Z_{\mathrm{R},k}(1+r_{\mathrm{F}}(p))}{Z_{\mathrm{L},k}Z_{\mathrm{R},k}P_{\mathrm{F}}(p)}Z_{\phi , k}\frac{\partial}{\partial p^2}P_{\mathrm{B}}(p)\frac{1}{(Z_{\phi , k}P_{\mathrm{B}}(p)+U_{k}')^2} \Big],
\end{eqnarray*}
respectively. Using the optimized regulator the result read in terms of dimensionless quantities
\begin{eqnarray}
\eta_{\mathrm{L}}&=&\frac{8v_d}{d}h_k^2[ m_{12}^{(\mathrm{FB})d}(\tilde{\rho}h_k^2, u_{k}'+2\tilde{\rho} u_{k}'')+m_{12}^{(\mathrm{FB})d}(\tilde{\rho}h_k^2, u_{k}') ],
\end{eqnarray}
and
\begin{eqnarray}
\eta_{\mathrm{R}}&=&\frac{8v_d}{d}h_k^2[ m_{12}^{(\mathrm{FB})d}(\tilde{\rho}h_k^2, u_{k}'+2\tilde{\rho} u_{k}'')+m_{12}^{(\mathrm{FB})d}(\tilde{\rho}h_k^2, u_{k}') +2(N_{\mathrm{L}}-1)m_{12}^{(\mathrm{FB})d}(0, u_{k}')].
\end{eqnarray}
\end{widetext}

\end{document}